\documentclass[10pt, twocolumn]{IEEEtran}

\usepackage{graphicx}
\usepackage{subfigure}
\usepackage{amsfonts}
\usepackage{cite}
\usepackage{amsmath, amsthm, amssymb}
\usepackage{algorithm}
\usepackage{algorithmic}

\usepackage{color}

\newtheorem{theorem}{Theorem}

\newtheorem{lemma}{Lemma}

\newtheorem{corollary}{Corollary}

\usepackage{amssymb}
\usepackage{makeidx}
\usepackage{amsmath}
\usepackage{graphicx}
\usepackage{epsf}
\usepackage{epsfig}
\usepackage{ccaption}
\usepackage{array}
\usepackage{tabularx}
\usepackage{multirow}
\usepackage{epsfig}
\usepackage{cite}
\usepackage{color}
\usepackage{wrapfig}

\long\def\symbolfootnote[#1]#2{\begingroup%
\def\thefootnote{\fnsymbol{footnote}}\footnote[#1]{#2}\endgroup}

\begin{document}
\bibliographystyle{IEEEtran}

\title{Achievable and Crystallized Rate Regions of the Interference Channel
with Interference as Noise}

\author{
\IEEEauthorblockN{Mohamad Awad Charafeddine, Aydin Sezgin, Zhu Han,
and Arogyaswami Paulraj}

\thanks{Manuscript received March 20, 2011; revised September 10, 2011.  This work is partially supported by
NSF CNS-0953377, CNS-0905556, CNS-0910461, and ECCS-1028782. The material in
this paper was presented in part at Allerton Conference on Communication, Control, and
Computing, Allerton, USA, September 2007 \cite{Charafeddine}, and at the IEEE
International Conference on Communications, Dresden, Germany, June 2009
\cite{CharafeddineCrystallized}.

M. Charafeddine was with the Department of Electrical Engineering at Stanford
University, CA, USA, and he now works at ASSIA, Redwood City, CA, USA (e-mail:
mohamad@stanfordalumni.org).

A. Sezgin is with the  Department of Electrical Engineering and Information
Technology at Ruhr-University Bochum, Germany (e-mail: aydin.sezgin@rub.de).

Z. Han is with the Electrical and
Computer Engineering Department, University of Houston, Houston, TX, USA, and with the Department of Electronics
and Radio Engineering, Kyung Hee University, Yongin, Gyeonggi, South Korea
(e-mail: zhan2@mail.uh.edu).

A. Paulraj is with the Department of Electrical Engineering at Stanford
University, CA, USA (e-mail: apaulraj@stanford.edu).

Copyright \copyright 2011 IEEE. Personal use of this material is permitted.
However, permission to use this material for any other purposes must be obtained from
the IEEE by sending a request to pubs-permissions@ieee.org.
}

}

\maketitle

\thispagestyle{empty}

\begin{abstract}
The interference channel achievable rate region is presented when the
interference is treated as noise. The formulation starts with the $2-$user
channel, and then extends the results to the $n-$user case. The rate region is
found to be the convex hull of the union of $n$ power control rate regions,
where each power control rate region is upperbounded by a $(n-1)$-dimensional
hyper-surface characterized by having one of the transmitters transmitting at full power.
The convex hull operation lends itself to a time-sharing operation depending on
the convexity behavior of those hyper-surfaces. In order to know when to use
time-sharing rather than power control, the paper studies the hyper-surfaces
convexity behavior in details for the $2-$user channel with specific results
pertaining to the symmetric channel. It is observed that
most of the achievable rate region can be covered by using simple On/Off binary
power control in conjunction with time-sharing. The binary power control
creates several corner points in the $n-$dimensional space. The crystallized
rate region, named after its resulting crystal shape, is hence presented as the
time-sharing convex hull imposed onto those corner points; thereby offering a
viable new perspective of looking at the achievable rate region of the
interference channel.

\end{abstract}

\IEEEpeerreviewmaketitle



\section{Introduction}
One important communication model in wireless communication is the
interference channel, which is subject to intensive research nowadays. For
example, the model is relevant for cellular networks in which multiple base
stations transmit data to their respective subscribers and thereby causing
interference at the unintended receivers, and ad-hoc networks in which nodes
are active at the same moment in the same frequency band. For a better
understanding of the interference channel, it is crucial to know its capacity
region, i.e., the maximum set of all achievable rate points. It serves also as a
benchmark for the comparison of different schemes. Unfortunately, the capacity
region of the $2-$user interference channel has been an open problem for about
$30$ years \cite{Sato:2usersCh, Sato:degardedGaussian2users}.
Information-theoretic bounds through achievable rate regions have been
proposed, most famously with the Han-Kobayashi region \cite{Kobayashi:interf}.
The capacity of the Gaussian interference channel under strong interference has
been found in \cite{Sato:capacityWithStrongInterf, Carleial}. Recent results on
the $2-$user interference channel to within one bit of capacity have been shown
in \cite{EtkinTse:GaussianInterfChannelToOneBit}, where a simplified
Han-Kobayashi scheme was used in which the message is split in two parts, the
private part and the common part. The transmit signal is then a superposition of
those two signals. By a smart allocation of power between those two parts it
was shown that this scheme is asymptotically optimal using a new metric, which
is referred to as the generalized degrees of freedom. The generalization of the
obtained results to the $n$-user case is rather difficult. As such, for the
$n$-user case mainly the capacity slope as a function of the
Signal-to-Noise-Ratio (SNR) are known. It was shown
in~\cite{CadambeJafar08_interfAlign} that for very high SNR, the capacity can
be approximated by $C=\frac{n}{2}\log(SNR)+o\left(\log(SNR)\right)$, where the
second term vanishes by definition for extremely high SNR.

The aforementioned referenced literature focused mainly on the $2-$user
interference channel from an information-theoretic point of view with highly
sophisticated and thus quite complicated transmitters and receivers. There are
other works in literature that tackle the practical issues in order to improve
the performance of the interference channels. Power control is one element of
critical importance. In \cite{Yates}, a framework for the uplink power control
is constructed and iterative power control is proposed. Adaptive modulation and
coding (AMC) can be combined with power control to enhance the network
performance \cite{Yanikomeroglu}. For multiple channel (such as OFDM) and
multiple cell case, joint AMC and power control have been widely employed
\cite{Su}. Beamforming and spatial diversity can also be utilized when
communicating over the MIMO channel \cite{Han_beamforming}. Interference
avoidance \cite{Rose} has also attracted many recent attentions. Finally, many
distributed solutions are proposed \cite{Saraydar,Hicks,zhu_book3} with the
benefit of simple implementation or low data overhead.

In this paper, the achievable rate
region is discussed for the $n-$user interference channel when the interference
is treated as additive Gaussian white noise and no multi-user detection is
employed. Examples where we encounter the need to define such rate region are
found in multicell communications, in addition to mesh and sensor
networks where the preference is to use low-complexity transceivers.
It is also interesting to note that using the strategy of treating interference
as additional noise proves to be asymptotically optimal, i.e., it was shown in
\cite{EtkinTse:GaussianInterfChannelToOneBit} that treating interference as
noise is optimal, as long as the interference power in dB is lower than
half of the useful signal power given that the power to noise ratio is
asymptotically high. This result was extended to the non-asymptotic case
independently by three research groups
\cite{KhandaniMotahari,VeeravalliAnna,KramerShangChen}. The generalization to
the asymmetric case and the $n-$user interference channel is given
in~\cite{KhandaniMotahari,VeeravalliAnna,KramerShangChen} as well, where it was
shown that it is optimal to treat the interference as noise whenever a similar
(sufficient) condition holds. The references
\cite{KhandaniMotahari,VeeravalliAnna,KramerShangChen} considered only the
Gaussian interference channel, while the general discrete many-to-one and
one-to-many memoryless channels were investigated in
\cite{CadambeJafarDiscreteMTIN}. It was shown in \cite{CadambeJafarDiscreteMTIN} that
treating interference as noise is also optimal in the discrete memoryless
channel as long as the received signal at the interfered receiver is
stochastically degraded compared to the received signals of the other
receivers. The optimality of treating interference as noise for the multiple
antenna case has been considered by~\cite{VeeravalliAnnaMIMO, ShangChenPoor,
BandemerSezginPaulraj}.

This paper finds the achievable rate region for the $n-$user interference
channel as the convex hull of the union of $n$ rate regions formed via power
control, where each rate region is upperbounded by a hyper-surface of dimension
$n-1$ characterized by having one of the transmitters operating at full power.
Given that there is a convex hull operation imposed onto the hyper-surfaces, it is
important to know their convexity behavior in order to determine when
time-sharing should be applied. This is treated in details for the $2-$user
interference channel. As the convex hull operation lends itself naturally to a
time-sharing operation, and based on the convexity conditions found, the paper
discusses when a time-sharing strategy should be employed rather than pure power
control, and then presents specific results pertaining to the $2-$user symmetric channel.
It is observed that the achievable rate region can be practically
approximated by using simple On/Off binary power control in conjunction with time-sharing. 
The On/Off binary power control creates several corner points in the
$n-$dimensional rate region, and employing a convex hull time-sharing operation
on those points achieves what is denoted as a crystallized rate region.

The system setup is presented in section \ref{sec_SysSetup}. Section
\ref{sec_FrontierTwoUser} discusses the achievable rate region
for the $2-$user interference channel, and then generalizes the results to the
$n-$user case. Section
\ref{sec_2usersRegion} focuses on characterizing the $2-$user
rate region in terms of convexity or concavity and when time-sharing is optimal with specific results to the
symmetric channel. Section \ref{sec:Crystallized} introduces the concept of the
crystallized rate region where time-sharing and On/Off binary power
control are used. Finally, the conclusion is drawn in section
\ref{sec:conclusions}.

\begin{figure} [t]
\centering
\includegraphics[width=166pt,height=120pt]{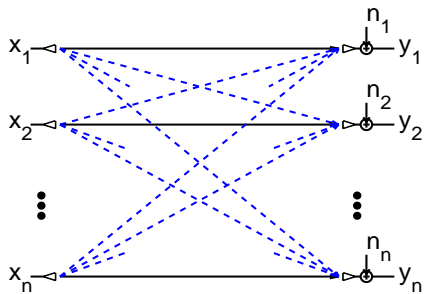}
\caption{$n-$user interference channel}
 \label{fig_n-user}
\end{figure}

\section{System Model}
\label{sec_SysSetup}
The $n-$user interference channel is presented
in Fig.\ref{fig_n-user} with $n$ transmitters and $n$ receivers.
The $i^{th}$ transmitter transmits its signal $x_i$ to the intended
$i^{th}$ receiver with power $P_i$. The receivers have independent
additive complex white Gaussian noise with zero mean and variance of
$\sigma_n^2$. Each transmitter is assumed to have a peak power
constraint of $P_{\max}$. Each transmitter has a single antenna and
communicates over a frequency flat channel. $g_{i,j}$ denotes the channel power gain received at
the $i^{th}$ receiver from the $j^{th}$ transmitter. Therefore, $g_{i,i}$ is the
channel gain of the $i^{th}$ desired signal, whereas $g_{i,j}$ with $j \neq i$
represents the interfering channel gain. {\bf P} is the transmit power vector of length $n$, where the
$i^{th}$ element $P_i$ denotes the transmit power of the $i^{th}$
transmitter. The interference is treated as additive noise throughout
this paper. $R_i$
denotes the maximum reliable rate of communication for the $i^{th}$
transmit-receive pair. Therefore, the achievable rate for the $i^{th}$ transmit-receive pair is written as:
\begin{align}
R_i({\bf P}) =
\log_2\left(1+\frac{g_{i,i}P_i}{\sigma_n^2+\sum_{j\neq
i}g_{i,j}P_j}\right). \label{Ci}
\end{align}
The next section finds the achievable rate region for such $n$ transmit-receive
pairs.

\section{Achievable Rate Region Frontiers for the Interference Channel}
\label{sec_FrontierTwoUser}

First, the section considers the $2-$user interference channel. The rate region
problem is analyzed by formulating its underlying nonconvex power control
problem, for which we find a closed form analytical solution. The rate region is
described by finding the maximum possible data rates achievable when each user
is subject to a maximum transmit power constraint. This
section then introduces the $3-$user case to study the effect of adding a new
dimension; and finally, by induction, the result is generalized for the $n-$user
case.

\subsection{$2$-user Achievable Rate Region Frontiers}

In the case of the $2-$user interference channel,
Eq.~(\ref{Ci}) can be expressed as a function of $P_1$ and $P_2$ as
$R_i(P_1,P_2),~i=1,2$. For notational brevity, the channel
gains are normalized by the noise variance, specifically:
$
a=g_{1,1}/\sigma_n^2$, $b=g_{1,2}/\sigma_n^2$,
$c=g_{2,2}/\sigma_n^2$, and $d=g_{2,1}/\sigma_n^2$.
$R_1$ and $R_2$ can therefore be written as:
\begin{align}
\begin{array}{c}
R_1(P_1,P_2) = \log_2\left(1+\frac{\displaystyle
aP_1}{\displaystyle 1+bP_2}\right),\\ R_2(P_1,P_2) =
\log_2\left(1+\frac{\displaystyle cP_2}{\displaystyle 1+dP_1}\right).
\end{array}
\label{R1P1P2orig}
\end{align}

For notational brevity, $\Phi(p_1,p_2)$ denotes a point in the rate region
marked by having $P_1=p_1$ and $P_2=p_2$. Effectively, the $x-$coordinate of
$\Phi(p_1,p_2)$ is $R_1(p_1,p_2)=r_1$, and the
$y-$coordinate of $\Phi(p_1,p_2)$ is $R_2(p_1,p_2)=r_2$.
The first objective is to find the achievable rate region frontiers of
Eq.~(\ref{R1P1P2orig}) through power control
of $P_1$ and $P_2$, where each transmitter is subject to the maximum
power constraint of $P_{\max}$. The frontier herein denotes the line (or
generally, the $(n-1)$-dimensional surface for the $n-$user channel) which
traces the rate region via power control.

\subsection{Rate Region Frontiers Formulation}
\label{FrontierFormulation} 
\begin{figure}[t]
\centering
\includegraphics[width=.5\textwidth]{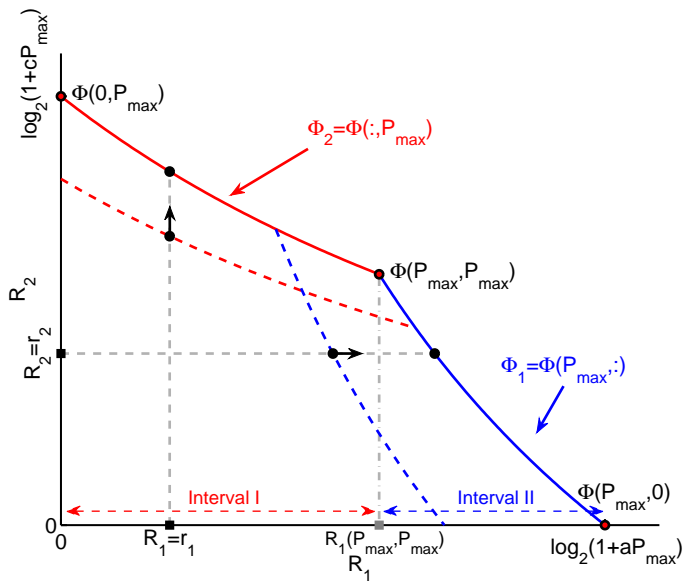}
\caption{$2-$user power-control rate region}
 \label{fig_2user_rateRegionFormulationMain}
\end{figure}

The rate region frontier can be traced by
setting $R_1$ to a certain value $r_1$, and then by sweeping $r_1$ over its
full possible range from $0$ to $R_1(P_{\max},0)$ while finding
the maximum $R_2$ value that can be achieved for each $r_1$. From
Eq.~(\ref{R1P1P2orig}), $R_1$ is monotonically increasing in $P_1$ and monotonically decreasing in $P_2$, thus point $R_1(P_{\max},0)$ corresponds to
point $\Phi(P_{\max},0)$ on the $x-$axis in Fig.
\ref{fig_2user_rateRegionFormulationMain}, representing the maximum
value $R_1$ can attain. Similarly for the $y-$axis, the maximum value
that $R_2$ can attain is $R_2(0,P_{\max})$, alternatively corresponding to
point $\Phi(0,P_{\max})$. Those points represent the cases in which one of the
users is silent, while the other is transmitting at full power.
Similarly, point $\Phi(P_{\max},P_{\max})$ has the
coordinates of $R_1(P_{\max},P_{\max})$ and $R_2(P_{\max},P_{\max})$.
	
\begin{figure}[t]
\centering
\includegraphics[width=.45\textwidth]{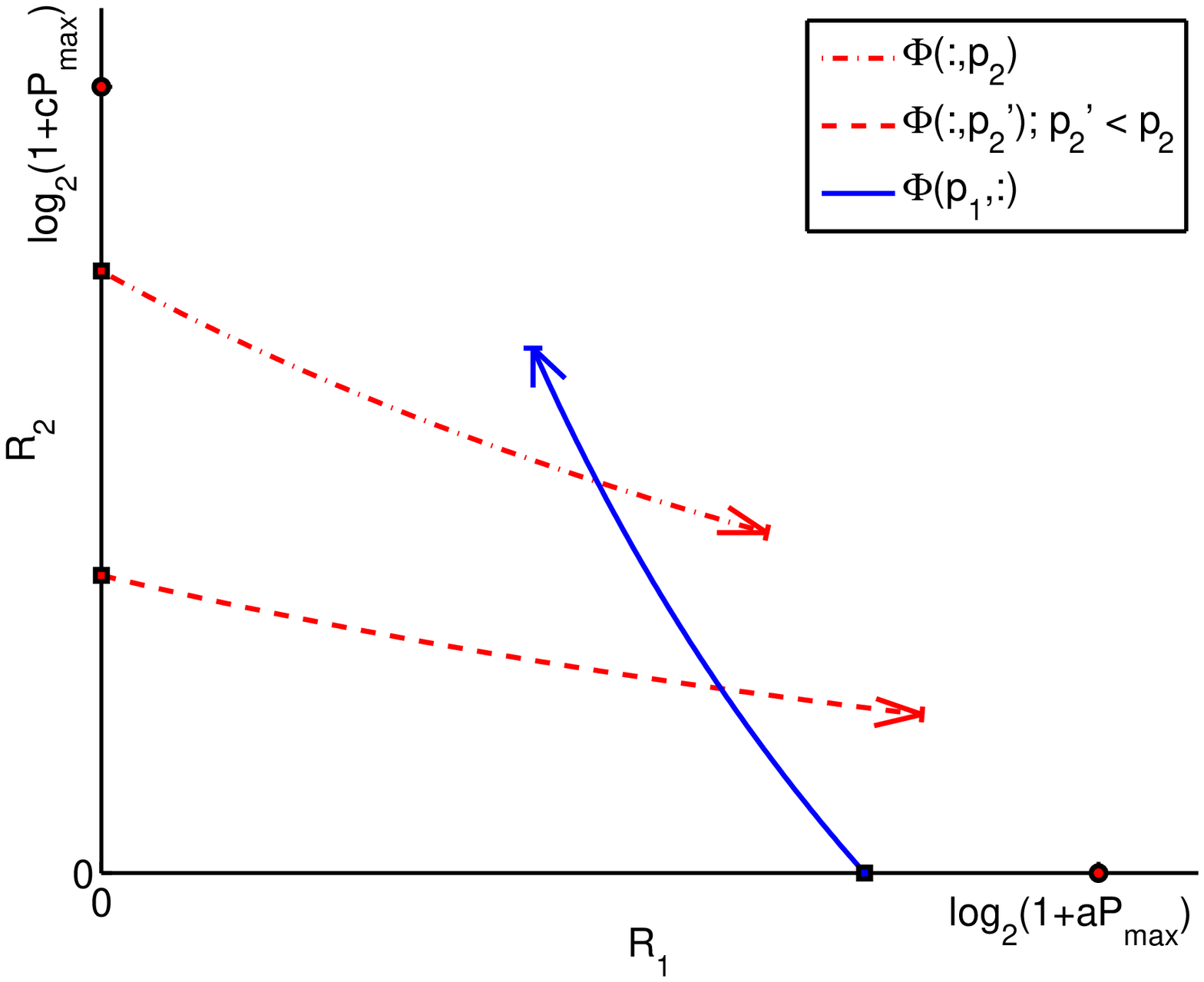}
\includegraphics[width=.45\textwidth]{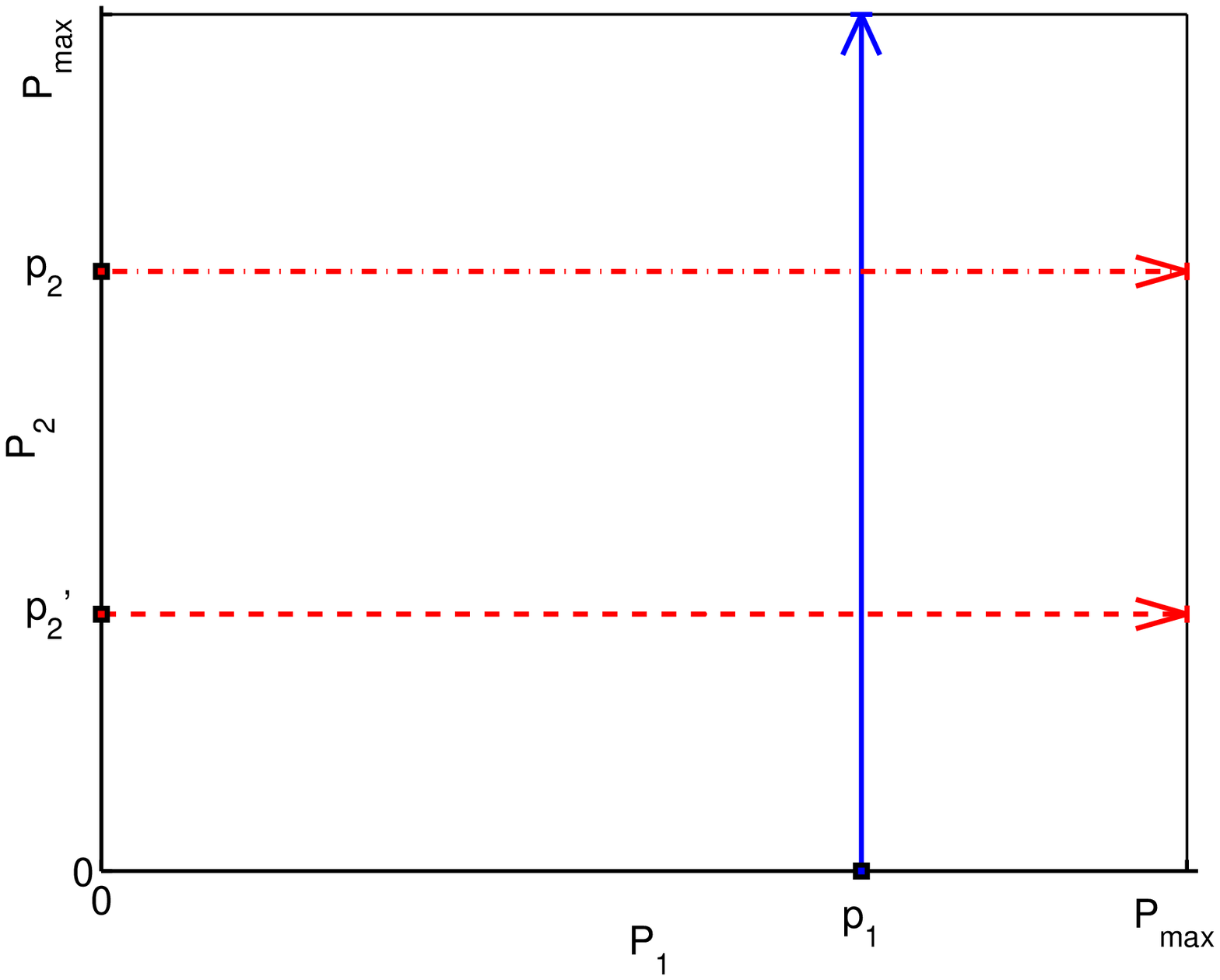}
\caption{potential lines illustration in rate region and power region}
 \label{fig_2user_potential_lines_illustration}
\end{figure}

Hence, for a constant rate $R_1=r_1$,
\begin{align}
R_1(P_1,P_2)=r_1=\log_2\left(1+\frac{aP_1}{1+bP_2}\right).
\label{C1P1P2}
\end{align}
Therefore, the relation between $P_1$ and $P_2$ is obtained as
follows:
\begin{align}
P_1=\frac{1}{a}(1+bP_2)(2^{r_1}-1). \label{P1P2relation}
\end{align}
From Eq.~(\ref{P1P2relation}), for a constant $R_1(P_1,P_2)=r_1$, $R_2(P_1,P_2)$
can now be written as a function of one parameter as $R_2(P_2)$, for $R_1=r_1$,
specifically:
\begin{align}
R_2(P_2) =
\log_2\left(1+\frac{\displaystyle cP_2}{\displaystyle 1+\frac{\displaystyle
d}{\displaystyle a}(1+bP_2)(2^{r_1}-1)}\right).
\label{C2P2}
\end{align}
It is important to analyze the behavior of $R_2(P_2)$ in terms of
$P_2$. 
This is presented in the following lemma: 
\begin{lemma}
Setting $R_1$ at a constant rate, $R_1(P_1,P_2)=r_1$, $R_2(P_2)$ is a
monotonically increasing function in $P_2$. \label{lemma_mono}
\end{lemma}
\begin{IEEEproof}
The proof is provided in Appendix~\ref{proof_lemma_mono}.
\end{IEEEproof}
Using this lemma, the following corollary of uniqueness property is obtained:
\begin{corollary}
For every rate tuple $(r_1^*,r_2^*)$, there is a unique power
tuple $(p_1^*,p_2^*)$.
\end{corollary}
\begin{IEEEproof} i) a direct implication of monotonicity in Eq.~(\ref{C2P2})
is that if $R_2$ is equal to a constant $r_2^*$ at the rate of $R_1=r_1$, then
there is a unique $p_2^*$ that achieves $r_2^*$, ii) when $p_2^*$ is determined,
then $P_1=p_1^*$ is uniquely defined from Eq.~(\ref{P1P2relation}), iii) from
$p_1^*$ and $p_2^*$, $R_1$ is uniquely defined as $R_1=r_1=r_1^*$ from
Eq.~(\ref{C1P1P2}). Thus, $p_1^*$ and $p_2^*$ uniquely define a point in
the rate region with coordinates $r_1^*$ and $r_2^*$.
\end{IEEEproof}
In other words, any point in the rate region is achieved solely by a
unique power tuple. This leads to what we denote by {\em potential lines}
$\Phi$ in the rate region, which are formed by holding one 
power dimension constant to a certain value and sweeping the other
power dimension over its full range. In that regard, to describe a potential
line marked by having $P_1$ held at a constant power $P_{\mbox{\small cst}}$, we
use the following notation $\Phi(P_{\mbox{\small cst}},:)$ to be equivalent to 
$\Phi(P_{\mbox{\small cst}},p_2)$ where $P_1=P_{\mbox{\small cst}}$ and $0 \leq
p_2 \leq P_{\max}$. Based on the uniqueness property just discussed, we have the
following corollary (illustrated in
Fig.\ref{fig_2user_potential_lines_illustration}):
\begin{corollary}
\label{cor:potential}
Potential lines\footnote{The property in Corollary \ref{cor:potential} is the
reason for denoting these lines as {\em potential lines}, where the nomenclature
is borrowed from electromagnetics based on a similar property for equipotential
lines of an electric field \cite{inanBook}.}
 along one power dimension do not intersect, i.e., $\Phi(:,p_2)$
and $\Phi(:,p_2')$ do not intersect if $p_2 \neq p_2'$.
\end{corollary}

The problem of finding the power control rate region frontiers then simplifies
into finding the maximum value $R_2(P_2)$ that can be achieved for any value of
$R_1(P_1,P_2)=r_1$. Effectively, the formulation of the power control rate
region is:
\begin{align}
\begin{array}{ll}
\arg \mathop {\max}\limits_{P_2} & R_2(P_2) \label{argC2}\\
\mbox{subject to} & R_1(P_1,P_2) = r_1,\\
& P_i \leq P_{\max},~~~~~i=1,2.
\end{array}
\end{align}
$r_1$ is swept over the full range of $R_1$, i.e., $0\leq r_1 \leq
R_1(P_{\max},0)$. 
The power control optimization problem in Eq.~(\ref{argC2}) is not
straightforward as it is a nonconvex problem \cite{boyd:convex}. However, by
splitting the rate range of $R_1$ into two intervals: {\em Interval 1} for
$0\leq r_1 \leq R_1(P_{\max},P_{\max})$ and {\em Interval 2} for
$R_1(P_{\max},P_{\max}) \leq r_1 \leq R_1(P_{\max},0)$, we are able to find a
closed form analytical solution for each interval. The analysis of the optimization problem in Eq.~(\ref{argC2}) over these two intervals follows in the next
two subsections.

\subsection{$R_2$ Frontier for Interval 1: $0\leq r_1 \leq
R_1(P_{\max},P_{\max})$} \label{subsec_Ra} As Eq.~(\ref{C1P1P2}) is monotonically increasing in
$P_1$ and monotonically decreasing in $P_2$, $r_1$ can only exceed
$R_1(P_{\max},P_{\max})$ when $P_2$ is less than $P_{\max}$. Thus,
$P_2=P_{\max}$ is attainable only when $0\leq r_1 \leq R_1(P_{\max},P_{\max})$,
and $P_2$ needs to be less than $P_{\max}$ otherwise. From the proof provided for Lemma \ref{lemma_mono}, where
Eq.~(\ref{C2P2}) is proved to be monotonically increasing in $P_2$, and
for the following {\em Interval 1} range of $r_1$: $0\leq r_1 \leq
R_1(P_{\max},P_{\max})$, the solution to the optimization problem is:
\begin{align}
\arg \max_{P_2} R_2(P_2)=P_{\max}. \label{P2Pmax}
\end{align}
Therefore, in this range of $r_1$, using
Eq.~(\ref{P1P2relation}) and Eq.~(\ref{P2Pmax}), $R_2$ is expressed by a
function of $r_1$ as follows:
\begin{align}
R_2(r_1)=
\log_2\left(1+\frac{\displaystyle cP_{\max}}{\displaystyle
1+\frac{\displaystyle d}{\displaystyle a}(1+bP_{\max})(2^{r_1}-1)}\right).
\label{eqPhi2}
\end{align}
Over this {\em Interval 1} range of $r_1$, the relation found in
Eq.~(\ref{eqPhi2}) describes the expression governing the potential line $\Phi(:,P_{\max})$, in
which $P_2$ is held at constant maximum power $P_{\max}$ and $P_1$ sweeps its
full range from $0$ to $P_{\max}$. For brevity, the potential line
$\Phi(:,P_{\max})$ is denoted as $\Phi_2$, where the second power dimension,
$P_2$, is held at the maximum power. In this range of $r_1$, $\Phi_2$ defined by
Eq.~(\ref{eqPhi2}) represents a power control frontier of the rate region
as shown in Fig.\ref{fig_2user_rateRegionFormulationMain}.

\subsection{$R_2$ Frontier for Interval 2: $R_1(P_{\max},P_{\max}) \leq r_1 \leq
R_1(P_{\max},0)$} \label{subsec_Rb} Using symmetry of the previous result, for a
constant rate $R_2=r_2$, there is a linear relation between
$P_1$ and $P_2$. Thus, $R_1(P_1,P_2)$ can be written in function
of one parameter $P_1$ as follows:
\begin{align}
R_1(P_1) =
\log_2\left(1+\frac{\displaystyle aP_1}{\displaystyle 1+\frac{\displaystyle
b}{\displaystyle c}(1+dP_1)(2^{r_2}-1)}\right).
\end{align}
By symmetry of the result in Lemma \ref{lemma_mono}, $R_1(P_1)$
is monotonically increasing in $P_1$. Thus, by symmetry, for the
following range of $r_2$: $
0\leq r_2 \leq R_2(P_{\max},P_{\max}), \label{RtildeRange}
$ we have:
\begin{align}
\arg \max_{P_1} R_1(P_1)=P_{\max}.
\label{RtildaPmax} 
\end{align}
Basically, the value found in Eq.(\ref{RtildaPmax}) describes the frontier for
the following rate ranges of: $0\leq r_2 \leq R_2(P_{\max},P_{\max})$ and
$R_1(P_{\max},P_{\max}) \leq r_1 \leq R_1(P_{\max},0)$ -- similar to
the former subsection \ref{subsec_Ra} where $P_2=P_{\max}$ described the
frontier for the following rate ranges of: $0\leq r_1 \leq
R_1(P_{\max},P_{\max})$ and $R_2(P_{\max},P_{\max}) \leq r_2 \leq
R_2(P_{\max},0)$.

Therefore, the value of $R_1$ at the frontier is:
\begin{align}
R_1(P_{\max},P_2)=\log_2\left(1+\frac{aP_{\max}}{1+bP_2}\right).
\end{align}
Hence, for this {\em Interval 2} range of $r_1$:
$R_1(P_{\max},P_{\max}) \leq r_1 \leq R_1(P_{\max},0)$, the value of $P_2$ that
achieves the frontier follows as:
\begin{align}
P_2=\frac{1}{b}\left(\frac{aP_{\max}}{2^{r_1}-1}-1\right).
\label{P2range2}
\end{align}
So effectively, the value found in Eq.~(\ref{P2range2}) is the answer for
the optimization problem in Eq.~(\ref{argC2}) for this range of $R$. 
Explicitly, for $R_1(P_{\max},P_{\max})\leq r_1 \leq
R_1(P_{\max},0)$, we have:
\begin{align}
\arg \mathop {\max}\limits_{P_2} R_2(P_2) =
\frac{\displaystyle 1}{\displaystyle b}\left(
\frac{\displaystyle aP_{\max}}{\displaystyle 2^{r_1}-1}-1\right).
\end{align}
Therefore, in this range of $r_1$, using
Eq.~(\ref{P2range2}) and Eq.~(\ref{C2P2}), $R_2$ is expressed in
function of $r_1$ as follows:
\begin{align}
R_2(r_1)=
\log_2\left(1+\frac{\displaystyle
\frac{c}{b}\left(aP_{\max}-(2^{r_1}-1)\right)}{\displaystyle
\left(2^{r_1}-1\right)(1+dP_{\max})}\right). \label{eqPhi1}
\end{align}
The relation found in Eq.~(\ref{eqPhi1}) describes the
expression governing the potential line $\Phi(P_{\max},:)$, where $P_1$ is held
at a constant maximum power $P_{\max}$ and $P_2$ sweeps its full range from
$0$ to $P_{\max}$. Similarly for brevity, the potential line
$\Phi(P_{\max},:)$ is denoted as $\Phi_1$. In this range of $r_1$, $\Phi_1$ as
defined by Eq.~(\ref{eqPhi1}) represents a power control frontier of the rate
region as shown in Fig.\ref{fig_2user_rateRegionFormulationMain}.
\begin{figure}[t]
\centering
\includegraphics[width=.5\textwidth]{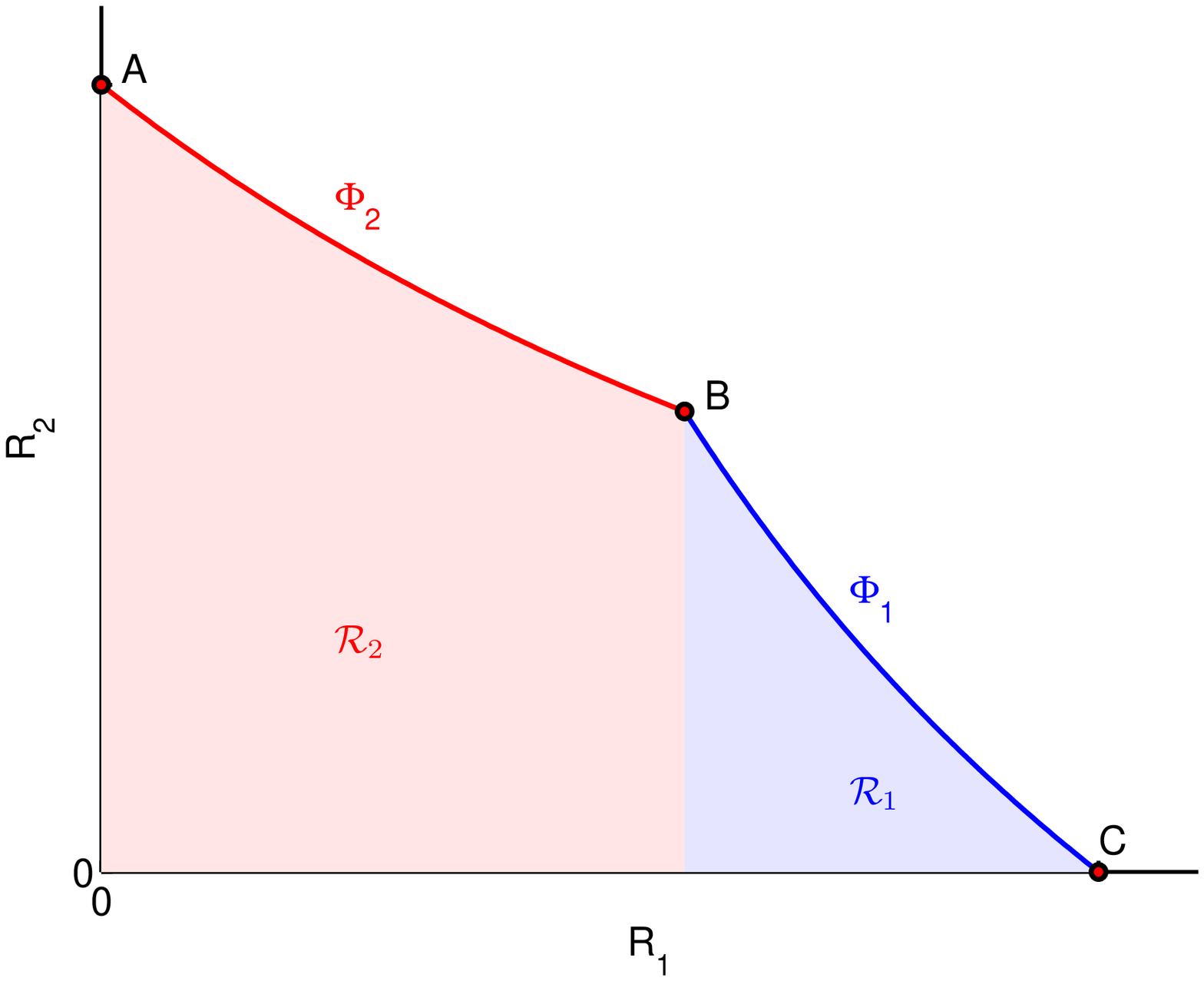}
\caption{rate regions ${\cal R}_1={\cal R}\{\Phi_1\}$ and ${\cal R}_2={\cal
R}\{ \Phi_2\}$}
 \label{fig_F1F2}
\end{figure}

\subsection{The $2-$User Achievable Rate Region}
This subsection consolidates the two earlier results to fully
describe the rate region frontiers.
\begin{itemize}
\item For {\em Interval 1} of $r_1$: $0\leq r_1 \leq R_1(P_{\max},P_{\max})$
\begin{align}
\arg \max_{P_2} R_2(P_2)=P_{\max}, \nonumber
\end{align}
and the power control frontier, $\Phi_2=\Phi(:,P_{\max})$,
is expressed as:
\begin{align}
R_2(r_1)= \log_2\left(1+\frac{\displaystyle cP_{\max}}{\displaystyle
1+\frac{\displaystyle d}{\displaystyle a}(1+bP_{\max})(2^{r_1}-1)}\right).
\label{F2}
\end{align}
Let ${\cal R}_2 = {\cal R} \{\Phi_2\}$ denotes the rate region outer-bounded by
$\Phi_2$ as shown in Fig. \ref{fig_F1F2}. 
\item For {\em Interval 2} of $r_1$: $R_1(P_{\max},P_{\max})\leq r_1 \leq
R_1(P_{\max},0) $
\begin{align}
\arg \max_{P_2}
R_2(P_2)=\frac{1}{b}\left(\frac{\displaystyle aP_{\max}}{\displaystyle
2^{r_1}-1}-1\right),
\nonumber
\end{align}
and the power control frontier, $\Phi_1=\Phi(P_{\max},:)$,
is expressed as:
\begin{align}
R_2(r_1)= \log_2\left(1+\frac{\displaystyle
\frac{c}{b}\left(aP_{\max}-(2^{r_1}-1)\right)}{\displaystyle
(2^{r_1}-1)(1+dP_{\max})}\right). \label{F1}
\end{align}
Similarly, let ${\cal R}_1 = {\cal R} \{ \Phi_1 \}$ denotes the rate region
outer-bounded by $\Phi_1$ as shown in Fig.\ref{fig_F1F2}.
\end{itemize}

In Fig.\ref{fig_F1F2}, point A denotes point $\Phi(0,P_{\max})$  (user $2$
transmitting solely at full power), point B denotes point
$\Phi(P_{\max},P_{\max})$ (both users are transmitting simultaneously at full
power), and point C denotes point $\Phi(P_{\max},0)$ (user $1$
transmitting solely at full power).

The rate region for a $2-$user interference channel achieved {\em through
power control} is obtained as:
\begin{align}
{\cal R}_1 \cup {\cal R}_2.
\end{align}
Finally, the $2-$user rate region, denoted as ${\cal R}$, is found as the convex
hull of the power control rate region. It is defined as:
\begin{align}
{\cal R}= \mbox{Convex Hull} \{ {\cal R}_1 \cup {\cal
R}_2 \}.
\end{align}

The treatment of
the achievable rate region for the $n-$user interference channel follows next.
It starts by considering a $3-$user interference channel to show the effect of
adding a new dimension, and then generalizes the result for the $n-$user case.
\begin{figure}[t]
\centering
\includegraphics[width=.5\textwidth]{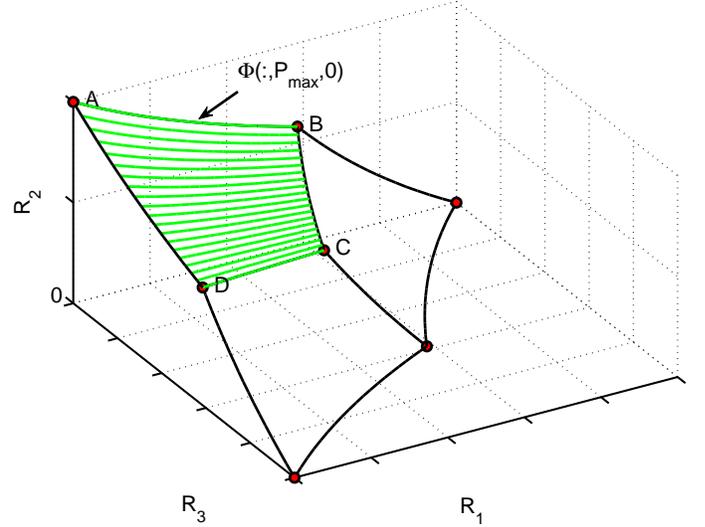}
\caption{3-user interference channel achievable rate region}
 \label{fig_3-userRateRegion}
\end{figure}

\subsection{$3-$User Example: Effect of Increasing $P_3$ from $0$ to $P_{\max}$}
The rate region for the $3-$user case is illustrated in
Fig.\ref{fig_3-userRateRegion}. The following notation of $\Phi(P_1,P_2,P_3)$
denotes a point in the rate region with coordinates of $[R_1(P_1,P_2,P_3)$,
$R_2(P_1,P_2,P_3)$, $R_3(P_1,P_2,P_3)]$. Accordingly, $\Phi(:,P_{\max},P_3)$
describes a line characterized by sweeping the transmit power $P_1$ of the first
transmitter from $0$ to
$P_{\max}$, with the second transmitter transmitting at $P_{\max}$ and the third transmitter
transmitting at a power value of $P_3$. Similarly, $R_i(:,P_{\max},:)$
represents a surface in the rate region marked by sweeping the full
range of $P_1$ and $P_3$, and holding $P_2$ at $P_{\max}$.

When $P_3=0$, the same setup and results that are described earlier in
this section applies. Thus, for the
rate range of $0\leq r_1 \leq R_1(P_{\max},P_{\max},0)$ and $0\leq
r_2 \leq R_2(0,P_{\max},0)$ and $R_3=0$, the frontier can be
described as $\Phi(:,P_{\max},0)$, which is the potential line from point A to
point B in Fig. \ref{fig_3-userRateRegion}. As $P_3$ increases, the goal is to
 describe its effect and how it is traced in the
rate region.

Revisiting Eq.~(\ref{Ci}), a fixed $P_3$ has the
effect of just an additive noise term in $R_1({\bf P})$ and
$R_2({\bf P})$. Hence, all the previous results in section
\ref{sec_FrontierTwoUser} are applicable for any value of $P_3$ in
describing the frontier for $R_1$ and $R_2$, since the effect of $P_3$
can be lumped in the noise term. Thus, for the range of $0\leq r_1
\leq R_1(P_{\max},P_{\max},P_3)$ and $0\leq r_2 \leq
R_2(0,P_{\max},P_3)$, where $P_3$ is constant, the frontier line on
$R_1$ and $R_2$ is $\Phi(:,P_{\max},P_3)$, i.e., characterized by
having $P_2=P_{\max}$. Consequently, the potential lines (or
surfaces) concept in the $3-$user case carries through.

Next, the frontier on $R_3$ is described. For each value of
$P_3$, $\Phi(:,P_{\max},P_3)$ traces one of the highlighted curves
in Fig. \ref{fig_3-userRateRegion}. For the collection of lines to
form a frontier, we want to prove that at each increasing value of
$P_3$ these non-intersecting potential lines monotonically increase in the $R_3$
dimension. This is evident from the relation between $R_3$ and $P_3$ in
Eq.~(\ref{Ci}). The maximum value of $R_3$ that can be achieved in this
case is when $P_3=P_{\max}$, i.e., $R_3(:,P_{\max},P_{\max})$.
Therefore, the highlighted frontier surface in Fig.
\ref{fig_3-userRateRegion} is the closed potential surface
$\Phi(:,P_{\max},:)$. The boundary contours of this surface are the
potential lines: $A\leftrightarrow B$, $B\leftrightarrow C$,
$C\leftrightarrow D$, and $D\leftrightarrow A$, defined as
$\Phi(:,P_{\max},0)$, $ \Phi(P_{\max},P_{\max},:)$,
$\Phi(:,P_{\max},P_{\max})$, and $\Phi(0,P_{\max},:)$, respectively.

By symmetry of interchanging $P_1$, $P_2$, and $P_3$, the $3-$user rate region
 is found via the convex hull onto the union of the regions bounded by these
 three surfaces: $\Phi(P_{\max},:,:)$, $\Phi(:,P_{\max},:)$, and $\Phi(:,:,P_{\max})$.
The rate region $\cal R$ is therefore expressed as:
$
{\cal R} = \mbox{Convex Hull} \{{\cal R}_1 \cup {\cal R}_2 \cup
{\cal R}_3\},
$
where ${\cal R}_i = {\cal R} \{\Phi_i\}$ is the region outer-bounded by
the potential surface $\Phi_i$, where
$\Phi_i=\Phi(\ldots,P_i=P_{\max},\ldots)$ is the surface characterized by having
$P_{\max}$ in the $i^{th}$ power position. (Note that the intersection of
potential surfaces is a potential line, as two of the dimensional inputs become
equal, i.e., $\Phi(P_{\max},P_{\max},:) \in {\cal R}_1$ and $
\Phi(P_{\max},P_{\max},:) \in {\cal R}_2$.)

\subsection{$n-$User Generalization}
The case for $n-$user generalization is done by induction. For
the $n^{th}$ added dimension to the existing $n-1$ dimensions
problem, the additional power effect of $P_n$ can be lumped in the
additive noise term of the existing expressions, and thus the
results for $R_1,\ldots,R_{n-1}$ hold and carry through. The potential
hyper-surfaces for fixed $P_n$ are non-intersecting and monotonically
increasing in $P_n$, and thus the maximum outer limit is reached with
$P_n=P_{\max}$ for the appropriate range of $R_1,\ldots,R_{n-1}$. Invoking
symmetry, we can generalize over all the rates ranges, therefore arriving to
the following theorem.

\begin{theorem}
\label{theorem_ratesRegion}
 The achievable rate region of the $n-$user interference channel by treating
 the interference as noise is:
\begin{align}
{\cal R}= \mbox{\textup{Convex Hull}} \{\cup_{i=1}^n {\cal R}_i\},
\end{align}
${\cal R}_i={\cal R}\{\Phi_i\}$, where $\Phi_i$ is a hyper-surface
frontier of $n-1$ dimensions, characterized by
holding the $i^{th}$ transmitter at full power.
\end{theorem}
Note that Theorem~\ref{theorem_ratesRegion} also holds for different
thermal noise levels or different maximum power levels.

\section{Convexity Characteristics of the Power Control Frontiers
For the $2-$User Interference Channel} \label{sec_2usersRegion}
This section focuses on the $2-$user interference channel and studies the
behavior of the power control frontiers, i.e., the potential lines $\Phi_1$ and
$\Phi_2$, in terms of convexity and concavity in order to determine when the
convex hull operation entails employing time-sharing. This happens whenever any
of the potential lines, or segment thereof, is convex, which enables higher
data rate to be achieved using time-sharing rather than using power control.
Furthermore, specific results pertaining to the symmetric channel are presented
at the end of this section.

The power control frontiers equations for the $2-$user case are:
\begin{itemize}
\item $\Phi_1$: $R_2(r_1)=
\log_2\left(1+\frac{\displaystyle
\frac{c}{b}\left(aP_{\max}-(2^{r_1}-1)\right)}{\displaystyle
\left(2^{r_1}-1\right)(1+dP_{\max})}\right)$,
\item $\Phi_2$: $R_2(r_1)=
\log_2\left(1+\frac{\displaystyle cP_{\max}}{\displaystyle
1+\frac{d}{a}(1+bP_{\max})(2^{r_1}-1)} \right).$
\end{itemize}
It is not clear when $\Phi_i$ is convex or concave, or whether it
can exhibit a non-stationary inflection point. The non-stationary inflection
point happens when the potential line has simultaneously a convex segment and a
concave segment. The convexity behavior is thus treated next in more
details.

\subsection{Convexity or Concavity of the Power Control Frontiers}
By using Eq.~(\ref{P1P2relation}) when
$P_2=P_{\max}$, the potential line $\Phi_2$ depends on $P_1$ through the
following relation of $r_1$ and $P_1$:
\begin{align}
P_1=\frac{1}{a}(1+bP_{\max})(2^{r_1}-1). \nonumber
\end{align}
Therefore, the second derivative of $\Phi_2$ with respect to
$r_1$ leads to the following expression in function of $P_1$:
\begin{align}
\frac{\partial^2 \Phi_2}{\partial r_1^2} =
(\alpha+adP_1)^2-(a-\alpha)(a -\alpha+acP_{\max}),
\end{align}
where $\alpha=d+dbP_{\max}$. 

If the potential line is
concave (i.e. $\frac{\partial^2 \Phi_2}{\partial r_1^2} \leq 0$)
then the enclosed region ${\cal R}\{\Phi_2\}$ is convex. The rate region is
defined to be convex when a straight line connecting any two points inside the
rate region is entirely enclosed in the rate region. In contrast, if the potential line is
not concave, i.e., if it is convex or exhibits a non-stationary inflection
point, then we describe its enclosed region as being concave; as in this
situation, the aforementioned definition of a convex region does not
hold. In summary, if $\Phi_i$ is concave, then ${\cal R}\{\Phi_i\}$ is convex,
and ${\cal R}\{\Phi_i\}$ is concave otherwise.

Let $\Re(\cdot)$ be the real operation, the {\em inflection
threshold} $Q_1$ is defined as:
\begin{align}
Q_1 = \frac{\Re(\sqrt{(a-\alpha)(a-\alpha+acP_{\max})})-\alpha}{ad},
\end{align}
where $Q_1$ was derived such that
$
\mbox{sign}\left(\frac{\partial^2 \Phi_2}{\partial r_1^2}\right)
= \mbox{sign}({P_1-Q_1}).
$
Therefore, it suffices to study convexity or concavity of potential
line $\Phi_2$ by examining the sign of $(P_1-Q_1)$. Note that the inflection
threshold $Q_1$ only depends on system parameters $a$, $b$, $c$,
$d$, and $P_{\max}$. The relation is nonlinear. By plugging in the respective
values, the convexity behavior is assessed, and it can be decided whether
time-sharing is needed. 

\begin{figure}[t]
\centering
\includegraphics[width=.5\textwidth]{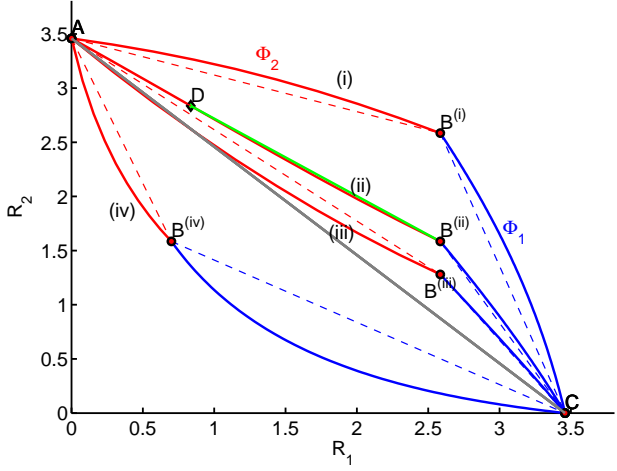}
\caption{$P_{\max}=1$; (i): concave $\Phi_2$ and $\Phi_1$
($[a,b;c,d]=[10,1;1,10]$), (ii): $\Phi_2$ with inflection point D, concave
$\Phi_1$ ($[10,1;4,10]$), (iii): convex $\Phi_2$, concave $\Phi_1$ ($[10,1;6,10]$),
(iv): convex $\Phi_2$ and $\Phi_1$ ($[10,15;4,10]$)}
\label{fig_multiInrConvexityCases}
\end{figure}

\begin{figure}[t]
\centering
\hspace{-.5cm}
\includegraphics[width=.25\textwidth]{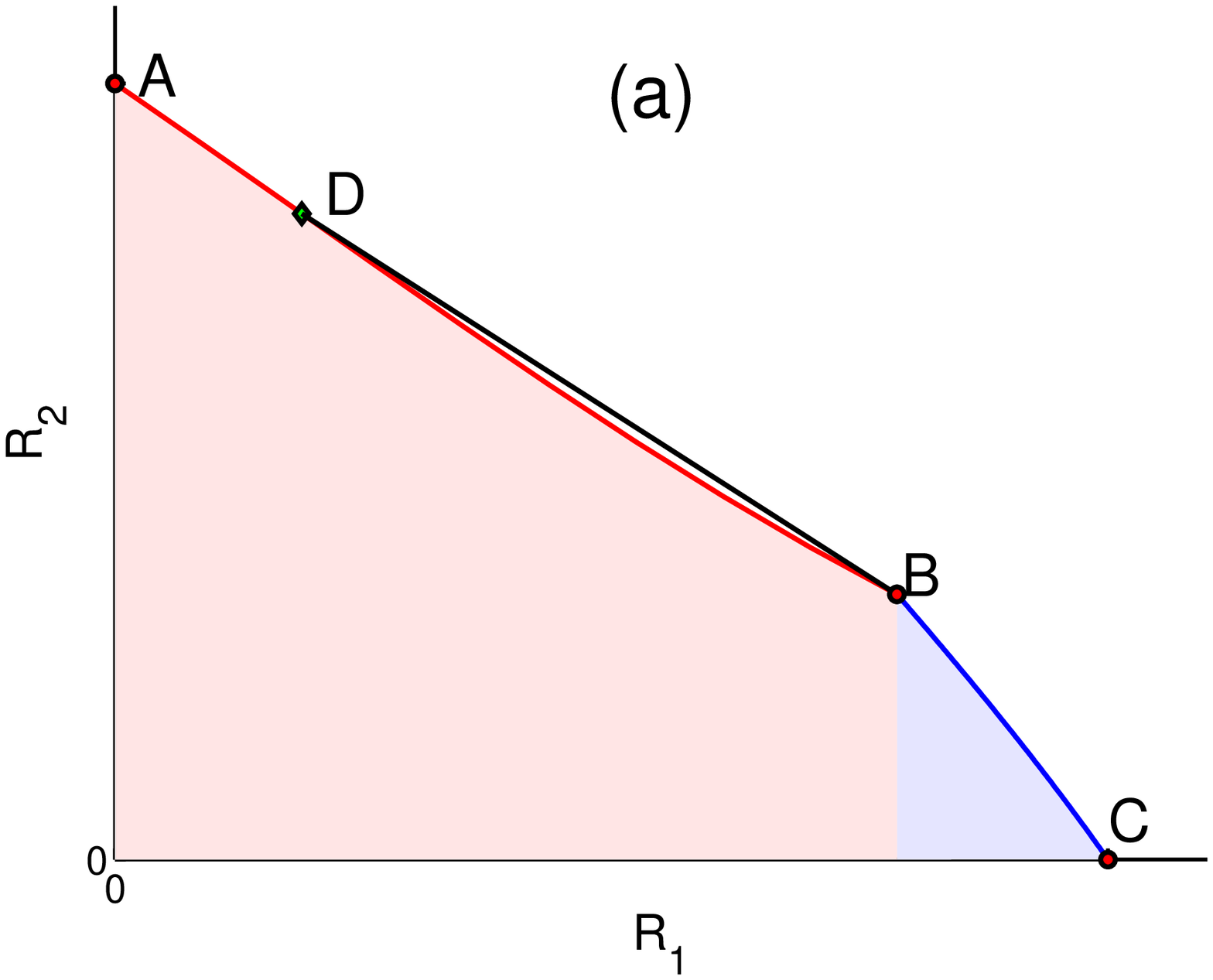}
\hspace{-.5cm}
\includegraphics[width=.25\textwidth]{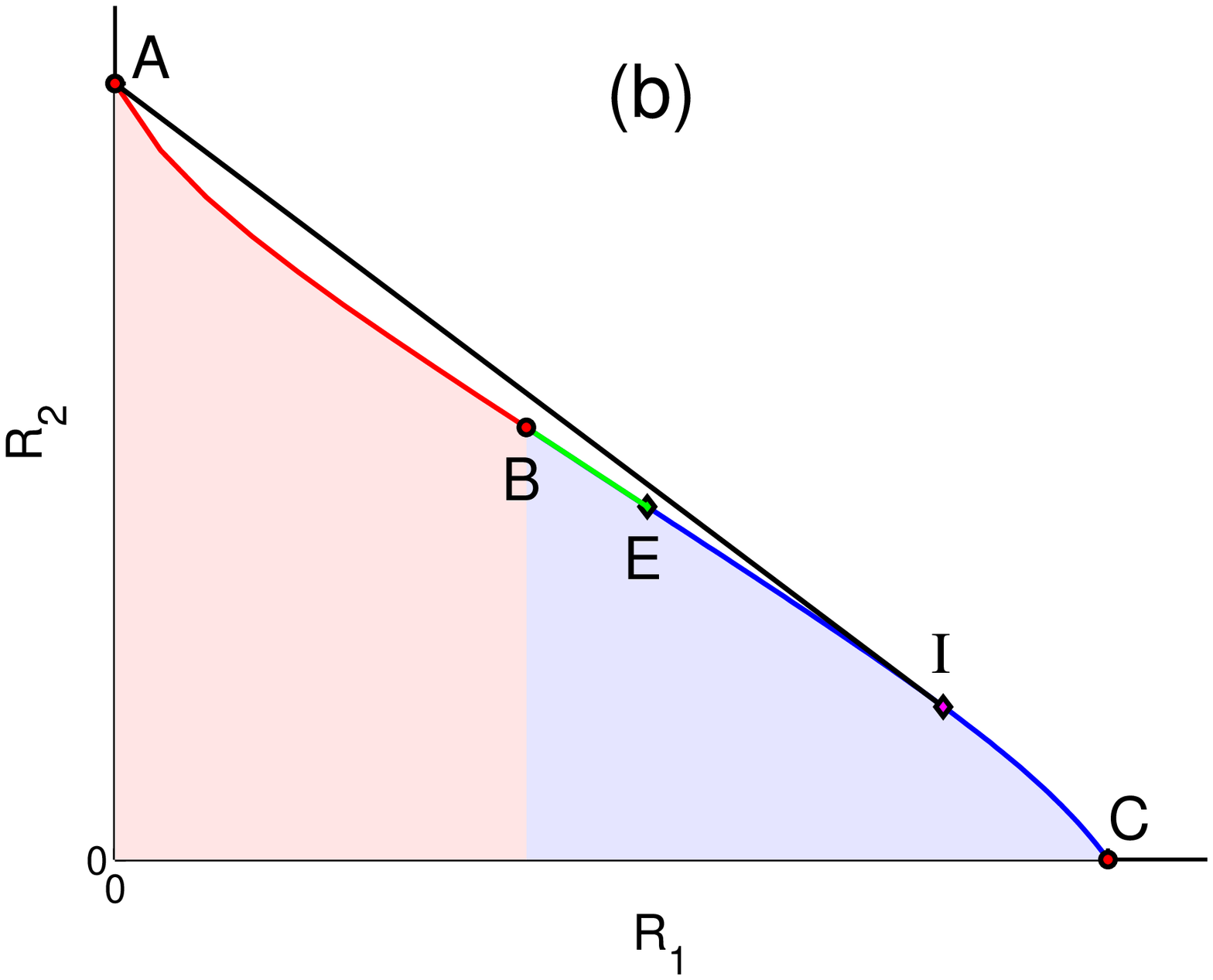}
\hspace{-.5cm}
\includegraphics[width=.25\textwidth]{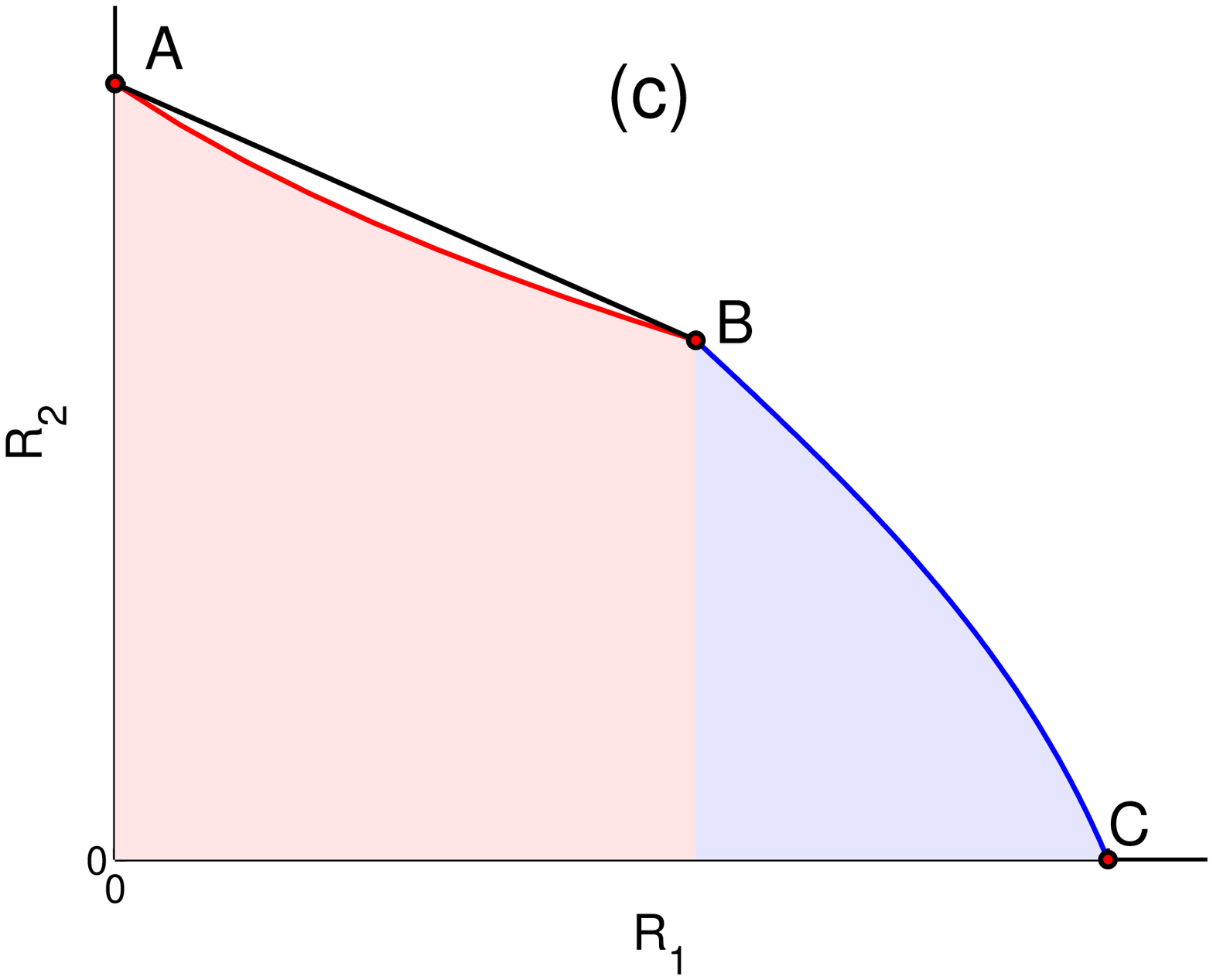}
\hspace{-.5cm}
\includegraphics[width=.25\textwidth]{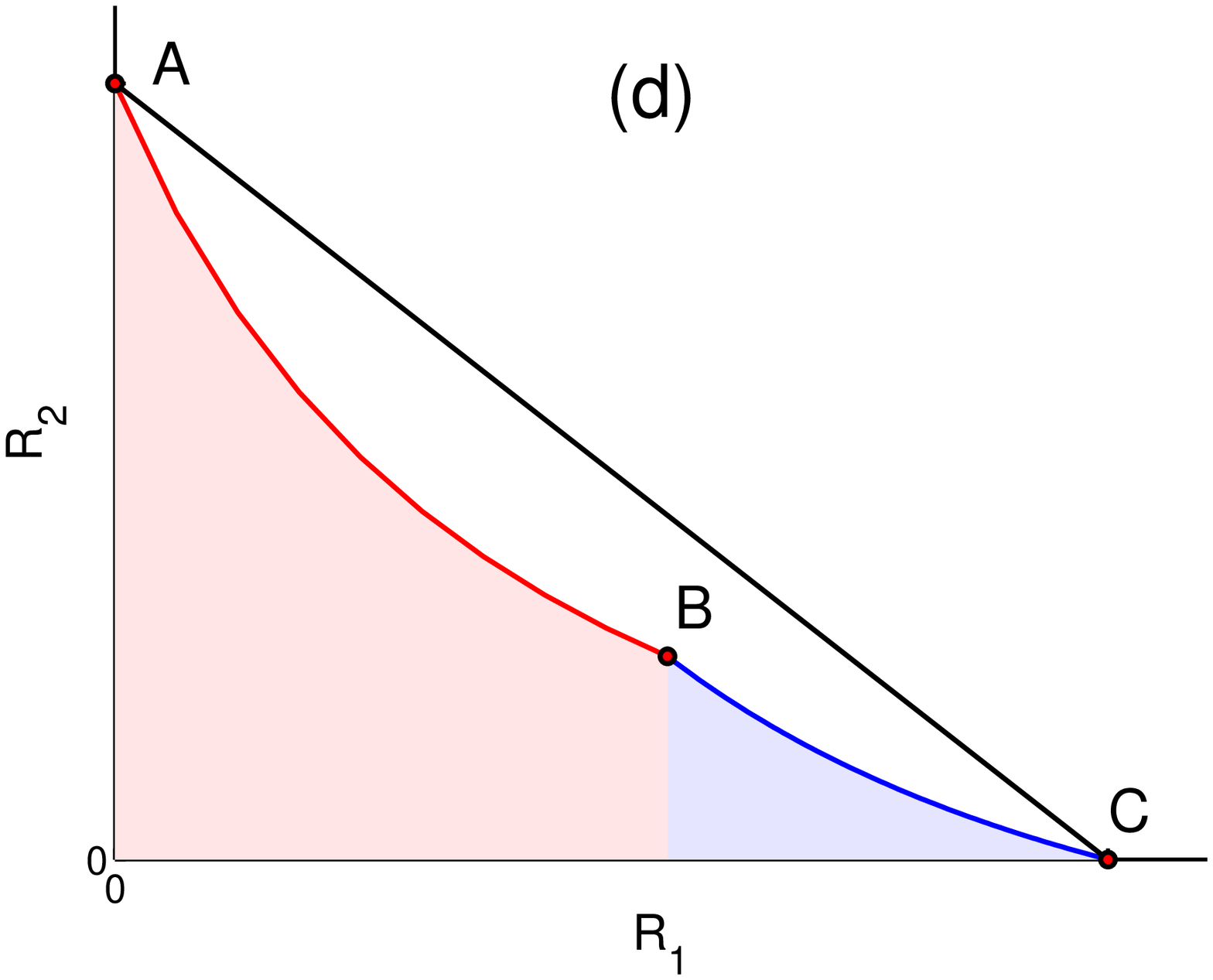}
\caption{types of time-sharing.}
\label{fig_types_of_time_sharing}
\end{figure}

Thus, $\Phi_2$ can exhibit the following convexity
behaviors:
\begin{itemize}
\item $Q_1 \geq P_{\max}$: then $P_1-Q_1 \leq 0$
 for all the range of $P_1$, thus $\Phi_2$ is concave. Operating via power
 control is optimal in leading the highest achievable data rate, and no
 time-sharing is needed. See $\Phi_2$ in case (i) in Fig.\ref{fig_multiInrConvexityCases}.
\item $0 < Q_1 < P_{\max}$: $\Phi_2$ exhibits a non-stationary
inflection point when $P_1=Q_1$ (see point D
in Fig.\ref{fig_multiInrConvexityCases} and
Fig.\ref{fig_types_of_time_sharing}). In this case:
    \begin{itemize}
     \item for $0 < P_1 \leq Q_1$: 
     line $\Phi(0:Q_1,P_{\max})$ is concave, i.e., the potential line segment
     $\Phi_{AD}$ is concave as in Fig.\ref{fig_multiInrConvexityCases} case
     (ii) and Fig.\ref{fig_types_of_time_sharing}~(a). Operating via power
     control to trace the segment $\Phi(0:Q_1,P{\max})$ is optimal.
     \item for $Q_1 \leq P_1 < P_{\max}$: line $\Phi(Q_1:P_{\max},P_{\max})$
     is convex, i.e., the potential line segment $\Phi_{DB^{(ii)}}$ is convex
     as in Fig.\ref{fig_multiInrConvexityCases} case (ii) and
     Fig.\ref{fig_types_of_time_sharing}~(a). Therefore, operating via the
     time-sharing segment between the inflection point D and point B is optimal.
\end{itemize}
\item $Q_1 \leq 0$: then $P_1-Q_1 \geq 0$
 for all the range of $P_1$, thus $\Phi_2$ is convex. See
 $\Phi_2$ in cases (iii) and (iv) in Fig.\ref{fig_multiInrConvexityCases}. Depending on the $a$, $b$,
 $c$, $d$, and $P_{\max}$ parameters, it is optimal to apply
 time-sharing with the following options:
 \begin{itemize}
   \item between point A and point B, see
   Fig.\ref{fig_types_of_time_sharing}~(c).
   \item between point A and a point on the concave segment of
   $\Phi_1$, see Fig.\ref{fig_types_of_time_sharing}~(b).
   \item between point A and point C, see
   Fig.\ref{fig_types_of_time_sharing}~(d). This is
   effectively Time Division Multiplexing (TDM), where each user transmits solely
   at any point of time. This form of dimension-orthogonality occurs when the
   interference is very strong rendering the cost too high for having
   simultaneous transmission. The upcoming subsection
   \ref{subsec:whenIsSimpleReuseOptimal} explores the optimality of operating via
   TDM, or equivalently, it explores when this cost is deemed too high.
 \end{itemize} 
\end{itemize}

By symmetry, the potential line $\Phi_1$ exhibits similar convexity
behavior: a) it is convex when $Q_2 \leq 0$, b) it is concave when
$Q_2 \geq P_{\max}$, and c) it exhibits a non-stationary inflection
point when $P_2=Q_2$. Hereby, with $\beta=(b+bdP_{\max})$, the inflection
threshold $Q_2$ is defined as:
\begin{align}
Q_2 = \frac{\Re(\sqrt{(c-\beta)(c-\beta+acP_{\max})})-\beta}{cb}.
\end{align}

Note that by virtue of how the inflection threshold is
situated with respect to power $P_i$, whenever any of the potential lines exhibit an
inflection point, the order of which segment is convex or concave is not
arbitrary. It always starts concave in the segment closer to the coordinate
axis. Explicitly, if $\Phi_2$ and $\Phi_1$ exhibit inflection points, then
tracing the rate region from left to right: $\Phi_2$ is concave then
transitions to convexity, and $\Phi_1$ is convex then transitions to concavity.

\subsection{When is TDM Optimal?}
\label{subsec:whenIsSimpleReuseOptimal}
Discounting the case when $\Phi_1$ or
$\Phi_2$ exhibit non-stationary inflection point for simplicity,
and focusing on the case when both potential
lines are convex (i.e., $Q_1 \leq 0$ and $Q_2 \leq 0$ ), it is important to know
when TDM is optimal. Under the aforementioned assumption, this translates to determine when time-sharing
between point A and point C is better than time-sharing through intermediate
point B. This is done by comparing the y-axis ordinate of point
B, $R_2(P_{\max},P_{\max})$, relative to the y-axis ordinate from
the straight line connecting points A and C at $r_1=R_1(P_{\max},P_{\max})$, denoted as
$R^{TS}_2(r_1)|_{r_1=R_1(P_{\max},P_{\max})}$, as
shown in Fig.\ref{fig_types_of_time_sharing}~(d). Namely, TDM is optimal
when $R^{TS}_2(r_1)|_{r_1=R_1(P_{\max},P_{\max})} \geq R_2(P_{\max},P_{\max})$;
that is:
\begin{align}
\begin{array}{c}
\frac{-\log_2(1+cP_{\max})}{\log_2(1+aP_{\max})}\log_2(1+\frac{aP_{\max}}{1+bP_{\max}})
+\log_2(1+cP_{\max}) \\
\geq \log_2(1+\frac{cP_{\max}}{1+dP_{\max}})
\end{array}
\nonumber
\end{align}

This leads to the following Lemma \ref{lemmaWhenTimeSharingIsOptimal}.

\begin{lemma}
\label{lemmaWhenTimeSharingIsOptimal}
Operating via TDM (i.e. one transmitter
solely transmitting at a certain time) is optimal in achieving
the rate region when
\begin{align}
\frac{(1+cP_{\max})(1+dP_{\max})}{1+cP_{\max}+dP_{\max}} \geq
\left(\frac{1+aP_{\max}+bP_{\max}}{1+bP_{\max}}\right)^\gamma
\label{condAC}
\end{align}
with $\gamma=\log_2(1+cP_{\max})/\log_2(1+aP_{\max})$.
\end{lemma}
Note that the condition found in Eq.~(\ref{condAC}) is a nonlinear relation
between the interference channel variables of $a, b, c, d$, and $P_{\max}$.
This motivates the following subsection \ref{subsec_sym2user} to treat the
$2-$user symmetrical channel.

\subsection{Symmetric $2-$User Interference Channel}
\label{subsec_sym2user}

This subsection treats the symmetric $2-$user channel, mainly analyzing the
expression in Eq.~(\ref{condAC}) in order to derive clear insights. For the
symmetric channel, $a=c$ and $b=d$, the expression in Eq.~(\ref{condAC})
simplifies, and leads to the following corollary on the TDM optimality condition:
\begin{corollary}
\label{cor:tdmOptimalSym} 
For the symmetric $2-$user interference channel, operating via TDM is optimal in
achieving the rate region when \begin{align}b \geq \frac{\displaystyle \sqrt{1+
aP_{\max}}}{\displaystyle P_{\max}}.
\label{b_threshold}
\end{align}
\end{corollary}
\begin{figure}[t]
\centering
\includegraphics[width=.5\textwidth]{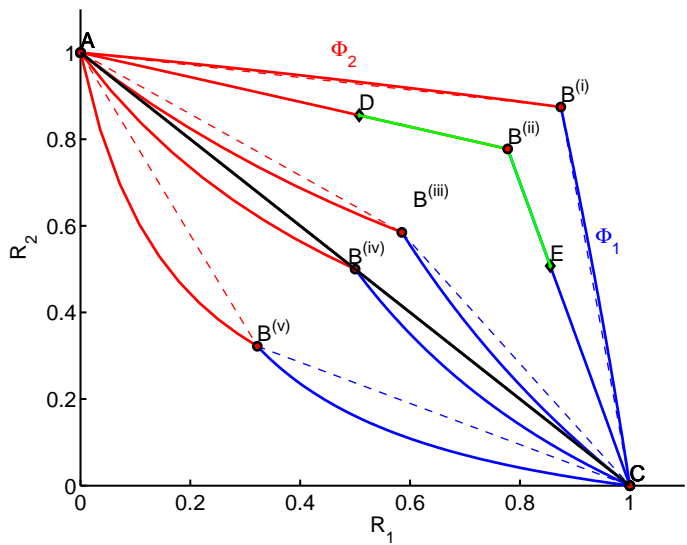}
\caption{symmetric $2-$user rate region: $P_{\max}=1$, $a=1$, $b^{(i)}=0.2$,
$b^{(ii)}=0.4$, $b^{(iii)}=1$, $b^{(iv)}=\sqrt{2}$, and $b^{(v)}=3$. The
threshold $b^*$ to switch to time-sharing is equal to $\sqrt{2}$
from Eq.~(\ref{b_threshold}).}
\label{fig_twoUserSym}
\end{figure}
In other words, when the interference is weak (i.e., $b$ is below the threshold
in Eq.~(\ref{b_threshold})), then it is best for both transmitters to transmit
at full power. When the interference increases and
exceeds the threshold in Eq.~(\ref{b_threshold}), then the TDM
scheme becomes optimal. In this scenario, the users can no longer share the same
resource, and thus they have to use it in an orthogonal fashion. An example is
illustrated in Fig.\ref{fig_twoUserSym}. The application of the usage of the different
types of time-sharing and the advantage over power control has been
discussed in \cite{CharafeddineCISS2009TwoSectors}, when it was applied to a
$2-$sector interference channel in a cellular setting.\\
\emph{Remark:} For high SNR (i.e. $aP_{\max} \gg 1$), Eq.~(\ref{b_threshold})
reduces to $bP_{\max} \geq \sqrt{aP_{\max}}$, which coincides with the results
in \cite{EtkinTse:GaussianInterfChannelToOneBit}; where this condition marks
the interference power threshold above which treating the interference as noise
is no longer optimal in the Degrees of Freedom sense and no longer within a gap
of $1$-bit from achieving capacity.

In addition, Appendix~\ref{App2} proves that the expression in
Eq.~(\ref{b_threshold}) is a sufficient condition for both frontiers $\Phi_1$
and $\Phi_2$ to be convex, i.e., $Q_1$ and $Q_2$ are always $\leq 0$; which is
the starting necessary condition of subsection
\ref{subsec:whenIsSimpleReuseOptimal} when treating the general asymmetric
channel. Therefore, when $b$ satisfies the expression in
Eq.~(\ref{b_threshold}), the frontiers potential lines are always convex and TDM is optimal.

The extension of Corollary \ref{cor:tdmOptimalSym} to the $n-$user symmetric interference channel is
provided in Appendix~\ref{App3}, leading to the following TDM optimality condition when:
\begin{align}
b \geq \left(\frac{\displaystyle 
aP_{\max}}{\displaystyle (1+aP_{\max})^{1/n}-1}-1
\right)\frac{\displaystyle 1}{\displaystyle (n-1)P_{\max}}.
\label{b_threshold_n}
\end{align}
\begin{figure}[t]
\centering
\includegraphics[width=.45\textwidth]{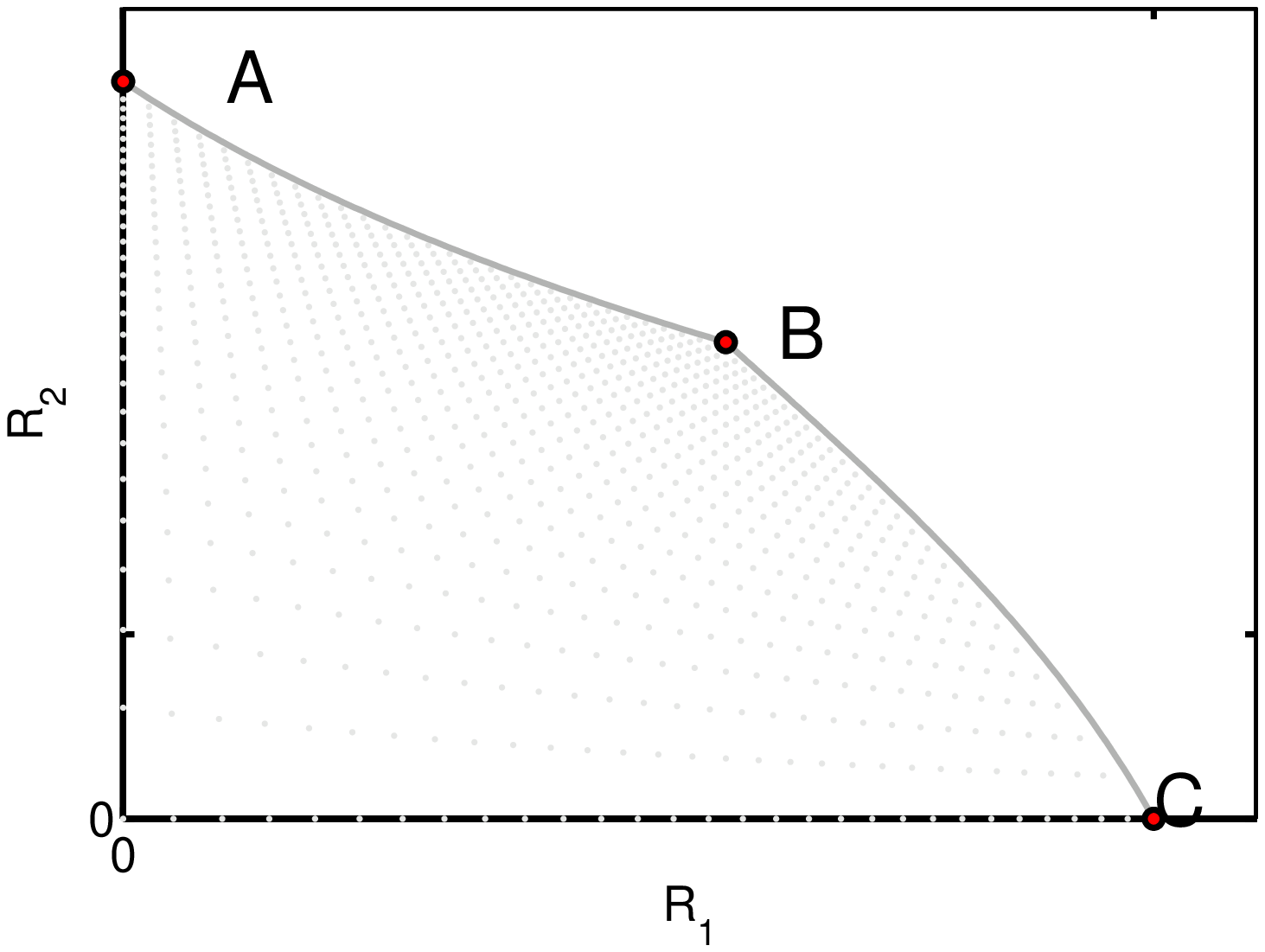}
\includegraphics[width=.45\textwidth]{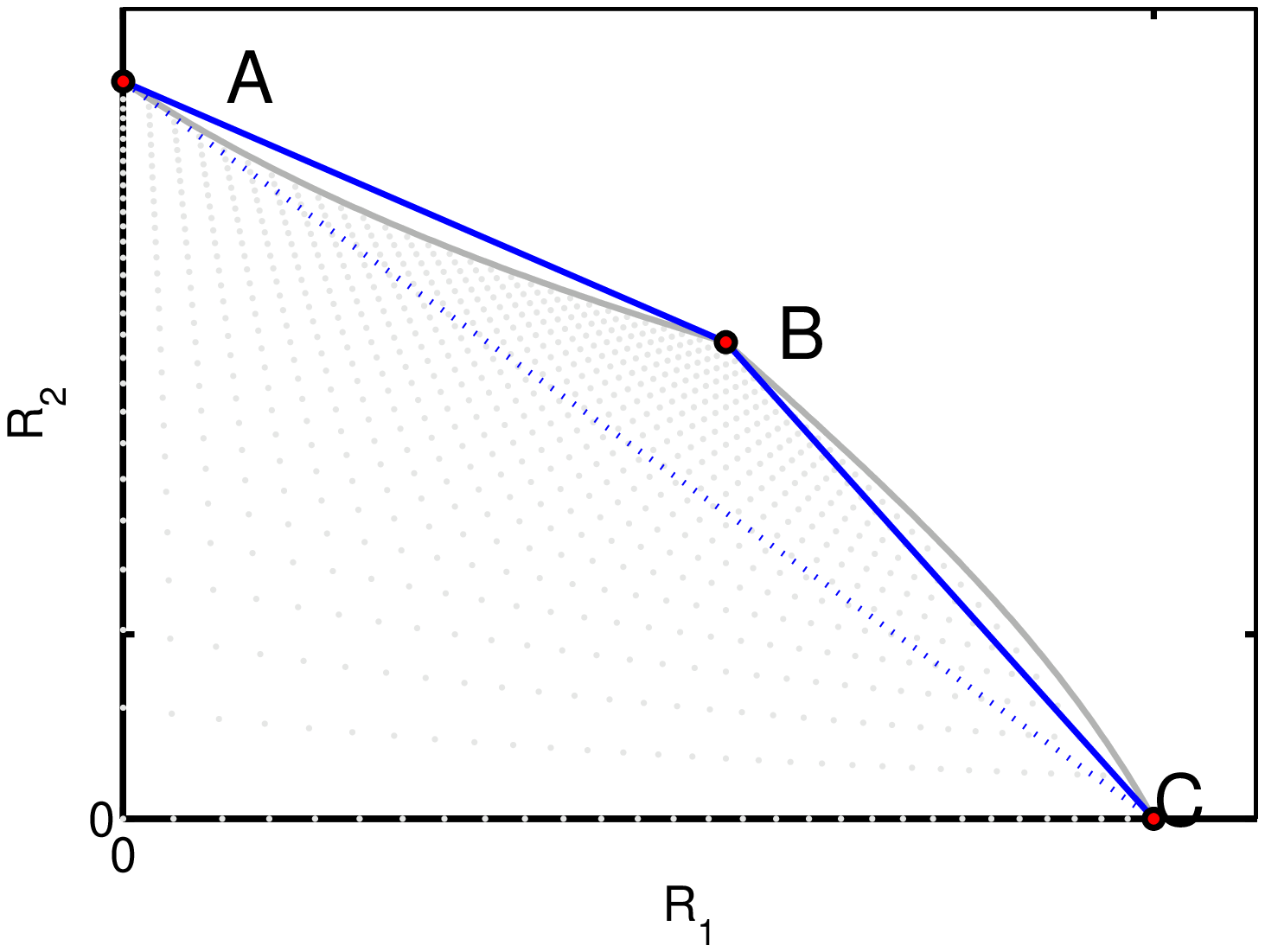}
\caption{$2-$user crystallized rate region: (a) rate region through power
control, (b) crystallized hull overlaid on top of the rate region}
\label{fig:2Drateregion}
\end{figure}
\section{Crystallized Rate Region}\label{sec:Crystallized}
Based on the discussion in the previous two sections, this section introduces a
novel approach into simplifying the rate region in the $n-$dimensional space by
having only an On/Off binary power control. This  consequently leads to $2^n-1$
{\em corner} points within the rate region. By forming a convex hull through
time-sharing between those corner points, it thereby leads to what we denote a
{\em crystallized} rate region \cite{CharafeddineCrystallized}. 
The concept of the crystallized rate region has since been extended to the MIMO
and the MIMO-OFDM channels \cite{DebbahCrystallizedMimoEurasip,
DebbahCrystallizedMimoOfdm}. This section
focuses on the crystallized rate region formulation and its evaluation.

\begin{figure}[t]
\centering
\includegraphics[width=.45\textwidth]{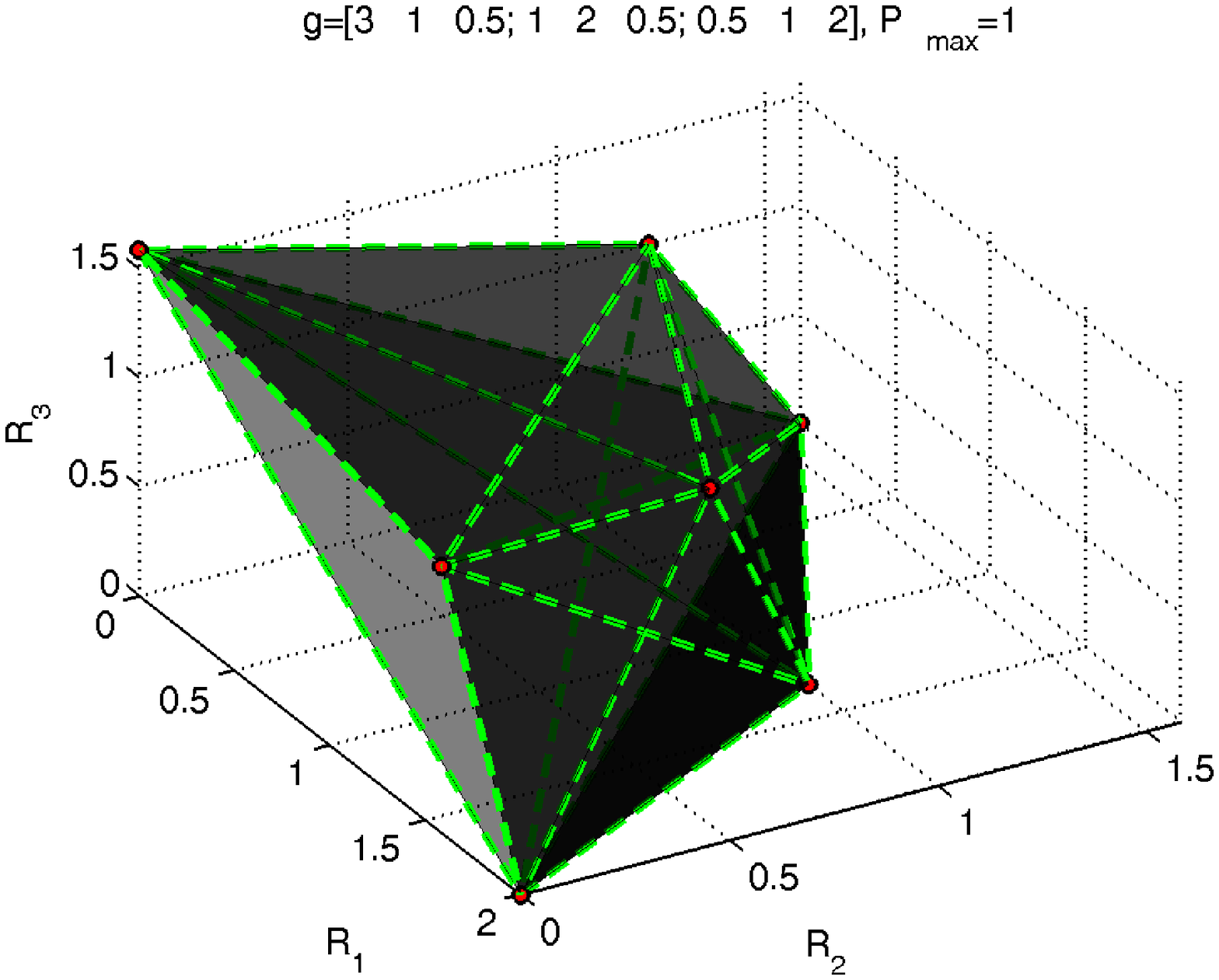}
\includegraphics[width=.45\textwidth]{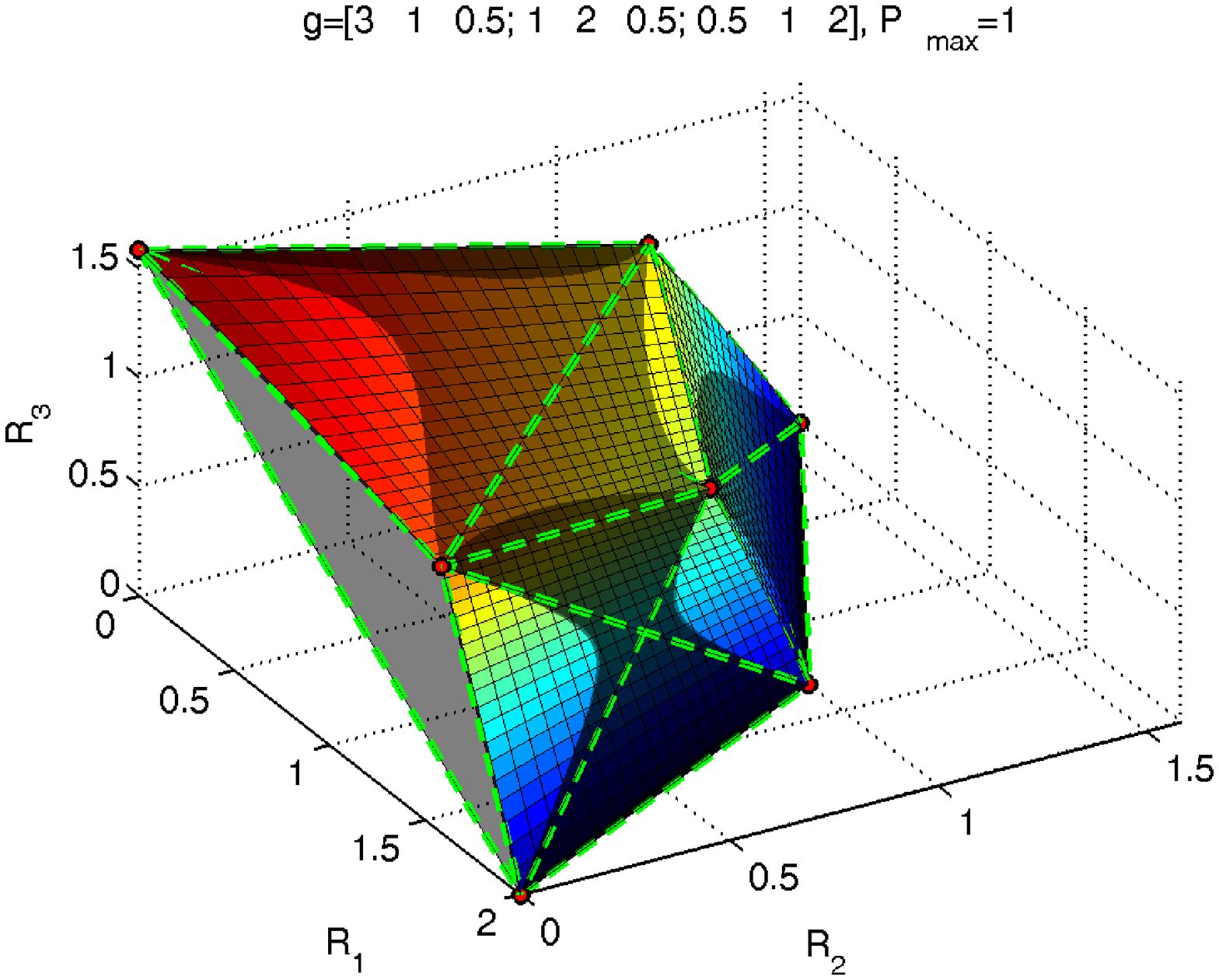}
\caption{3-user crystallized rate region: (a) time-sharing crystallized
hull, (b) crystallized hull overlaid on top of the power control rate region}
\label{fig:3users}
\end{figure}

As illustrated in Fig.\ref{fig:2Drateregion} for the $2-$user case, the
crystallized rate region is an approximation of the achievable rate region
formed by a convex hull of straight lines connecting points A, B, and C.
Those corner points are achieved through binary power control
with the transmitters employing either zero or full power. For the $2-$user
interference channel, there exist $3$ corner points; similarly for the $n-$user
case, there exist $2^n-1$ corner points. Note that in
Fig.\ref{fig:2Drateregion}, the time-sharing convex hull connects point A and
point C though point B; whereas if the interference is strong, the convex hull
is formed by time-sharing A and C only. Fig.\ref{fig:3users}
shows an example of the $3-$user case where Fig.\ref{fig:3users}~(a) is
the crystallized hull, and Fig.\ref{fig:3users}~(b) is the
crystallized hull overlaid on top of the power control rate region.
In the $2-$user dimension, the time-sharing convex hull is a set of
straight lines connecting two points. In the $3-$user
dimension, it is a set of polygon surfaces connecting three
points, see Fig.~\ref{fig:3users_polygon}. In the $n-$user dimension, each
polygon is formed by connecting $n$ points, and hence the number of such polygons is
$2^n-1 \choose n$.

\subsection{System Time-sharing Coefficients and New Rates Equations}
\label{subsec:thetas}

Recall that with power control, the achievable rates for the
$2-$user interference channel are given in Eq.~(\ref{R1P1P2orig}).
The paradigm with the crystallized rate region approach is different. Instead
of formulating the problem as a power control problem for finding $P_i$, the
crystallized rate region formulation focuses on finding the appropriate
time-sharing coefficients of the $2^n-1$ corner points. For the $2-$user case,
let {\boldmath $\theta$}$=[\theta_1, \theta_2, \theta_3]^T$,
$\sum_i\theta_i=1$, denote the {\em system} time-sharing coefficients vector of
respective corner points $\Phi(P_{\max},0)$ (user 1 transmitting solely with a
time-sharing coefficient $\theta_1$), $\Phi(0,P_{\max})$ (user 2 transmitting
solely with a time-sharing coefficient $\theta_2$), and $\Phi(P_{\max},
P_{\max})$ (both users transmitting with  a time-sharing coefficient
$\theta_3$). The reason for labeling {\boldmath $\theta$} a {\em system}
time-sharing coefficients vector is to emphasize the combinatorial element in
constructing the corner points, where the cardinality of $|${\boldmath $\theta$}$|=2^n-1$.

For $2-$user case, in contrast with Eq.~(\ref{R1P1P2orig}), the rates 
$R_1$ and $R_2$ for the crystallized region are:
\begin{align}
\begin{array}{l}
R_1(\mbox{\boldmath $\theta$})=\theta_1\log_2(1+aP_{\max})
+\theta_3\log_2\left(1+\frac{\displaystyle aP_{\max}} {\displaystyle
1+bP_{\max}}\right), \\
R_2(\mbox{\boldmath$\theta$})=\theta_2\log_2(1+cP_{\max})
+\theta_3\log_2\left(1+\frac{\displaystyle cP_{\max}}
{\displaystyle 1+dP_{\max}}\right).
\end{array}
\label{crystallizedRates}
\end{align}
\begin{figure}[t]
\centering
\includegraphics[width=.45\textwidth]{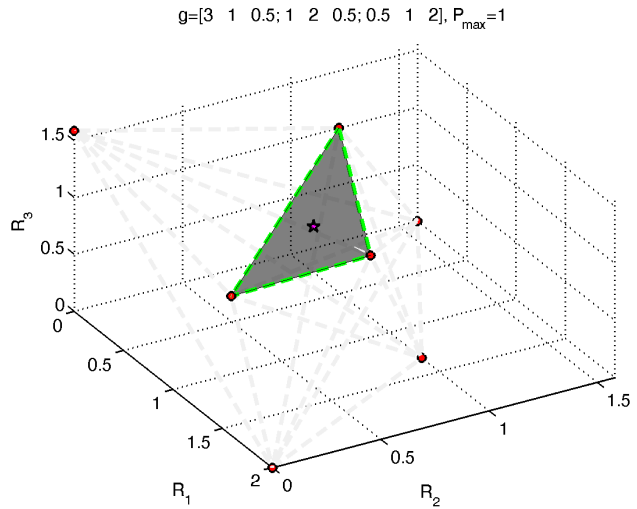}
\caption{a desired operating point in a time-sharing polygon}
\label{fig:3users_polygon}
\end{figure}
Any solution point on the crystallized hull lies somewhere on a
time-sharing line connecting two points for the $2-$user case; and similarly in
the $3-$user case, the solution point lies somewhere on a time-sharing
plane connecting three points (see Fig.\ref{fig:3users_polygon}), then by
induction we obtain the following corollary:
\begin{corollary}
\label{cor:max_n_nonzero}
For any solution point on the
$n-$user crystallized rate region, the system time-sharing vector {\boldmath
$\theta$} has at maximum $n$ nonzero coefficients out of its $2^n-1$ elements.
\end{corollary}
Corollary \ref{cor:max_n_nonzero} could also have been reached invoking the
Carath$\acute{\mbox{e}}$odory theorem
\cite{book:convexAnalysis}, where, in the paper's context, any point lying on the hyper-surface of dimension $n-1$ is
enclosed in a convex set of $n$ or fewer points inside the hyper-surface.
Therefore, if those points are the corner points, any solution point on the
$n-1$-dimensional hyper-surface will be the result of time-sharing at maximum
$n$ corner points.

Let {\boldmath$\alpha$}$^{(k)}$ denote the transmitters power mask that
characterizes the $k^{th}$ corner point, specifically:
\begin{align}
\begin{array}{c}
{\mbox{\boldmath$\alpha$}}^{(k)} =[\alpha_1^{(k)},\cdots,\alpha_i^{(k)},\cdots, \alpha_n^{(k)}]^T,
 \alpha_i^{(k)} \in \{0,1\}\\
i=1,\cdots,n \mbox{, where $i$ is the transmitter index,}\\
k=1,\cdots,2^n-1 \mbox{, where $k$ is the corner point index.}\\
\end{array}
\end{align}
$\alpha_i^{(k)}$ is the action taken by the $i^{th}$ user in characterizing the
$k^{th}$ corner point; the binary transmit action is either being silent or
transmitting at full power. Let {\boldmath$\alpha$}$_{-i}^{(k)}$ indicate the
interferers transmit power mask, which is derived from vector
{\boldmath$\alpha$}$^{(k)}$ by excluding the $i^{th}$ transmitter's action. The
generalization for the $n-$user crystallized rates equations now follows as:
\begin{align}
R_i(\mbox{\boldmath$\theta$})=\sum_{k=1}^{2^n-1}\theta_k\log_2
\left(1+\frac{\displaystyle \alpha_i^{(k)}g_{ii}P_{\max}}{\displaystyle 1+
{{\mbox{\boldmath$\alpha$}}_{-i}^{(k)}}^T{\mbox{\boldmath$g$}}_iP_{\max}}
\right),
\label{eq:generalizedCrystallizedRates}
\end{align}
where $g_{ii}$ is the desired channel gain from the $i^{th}$ transmitter to the
$i^{th}$ receiver normalized by the noise power; {\boldmath $g$}$_i$ is the
noise-normalized channel gains vector received at the $i^{th}$ receiver from all
the transmitters excluding the $i^{th}$ transmitter. The length of
each vector {\boldmath $\alpha$}$_{-i}^{(k)}$ and {\boldmath $g$}$_i$ is $n-1$.

\subsection{Evaluation of Crystallization}
In this section, we compare the crystallized rate region and the rate region
bounded by the power control potential lines. In effect, we are
evaluating how much gain or loss results from completely replacing
the traditional power control scheme (see
Eq.~(\ref{R1P1P2orig})) with the time-sharing scheme between the
corner points (see Eq.~(\ref{crystallizedRates})). For this purpose,
we consider the symmetric channel, where $a=1, P_{\max}=1$, and we increase
the interference $b$ to vary the signal to interference ratio $\mbox{SIR} =
a/b$. Two metrics are used as a measure: (a) the area of
the rate region, and (b) the maximum bit rate gap between the traditional
power control and the crystallized rate region scheme.

\begin{figure}[t]
\centering
\includegraphics[width=.45\textwidth]{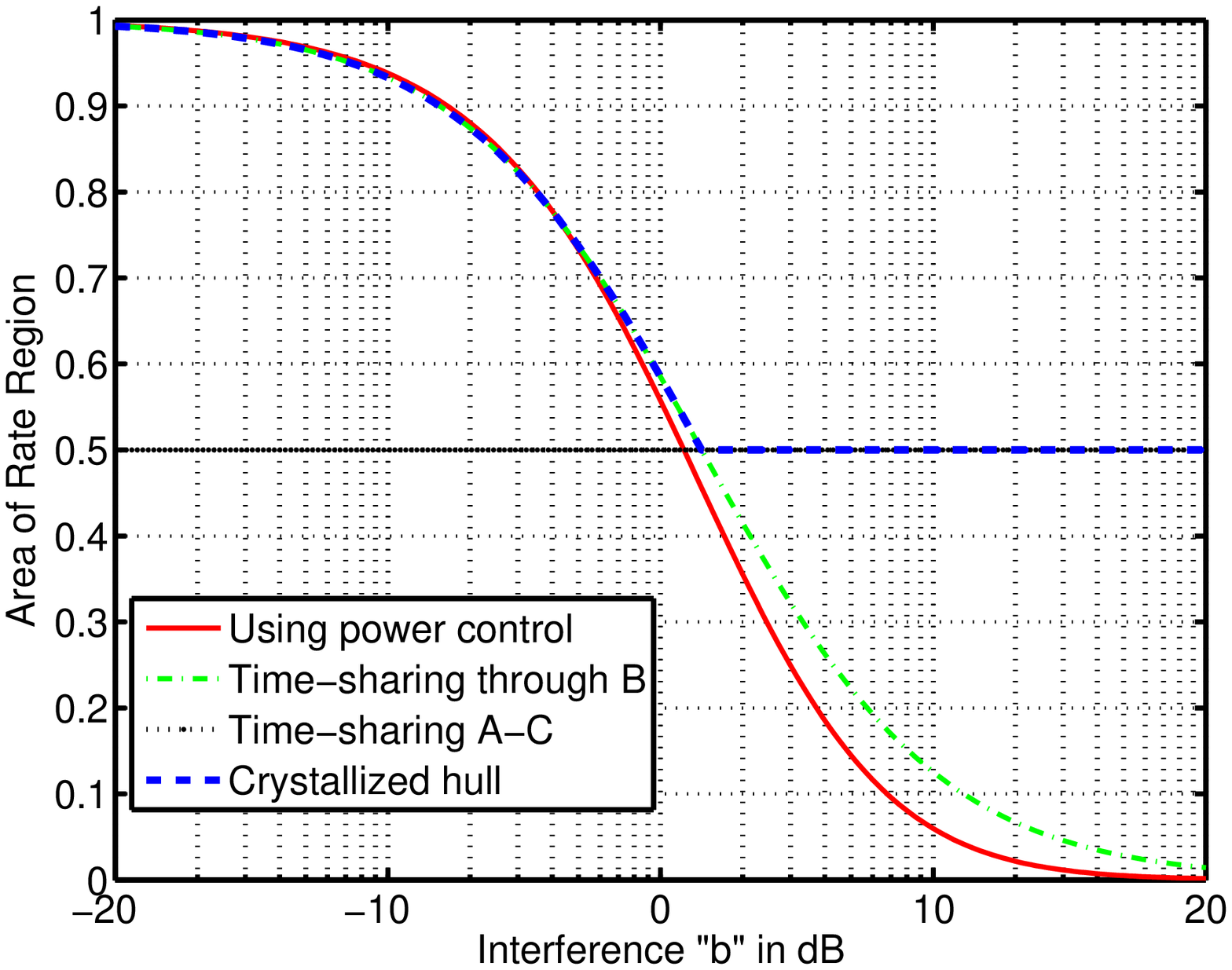}
\includegraphics[width=.45\textwidth]{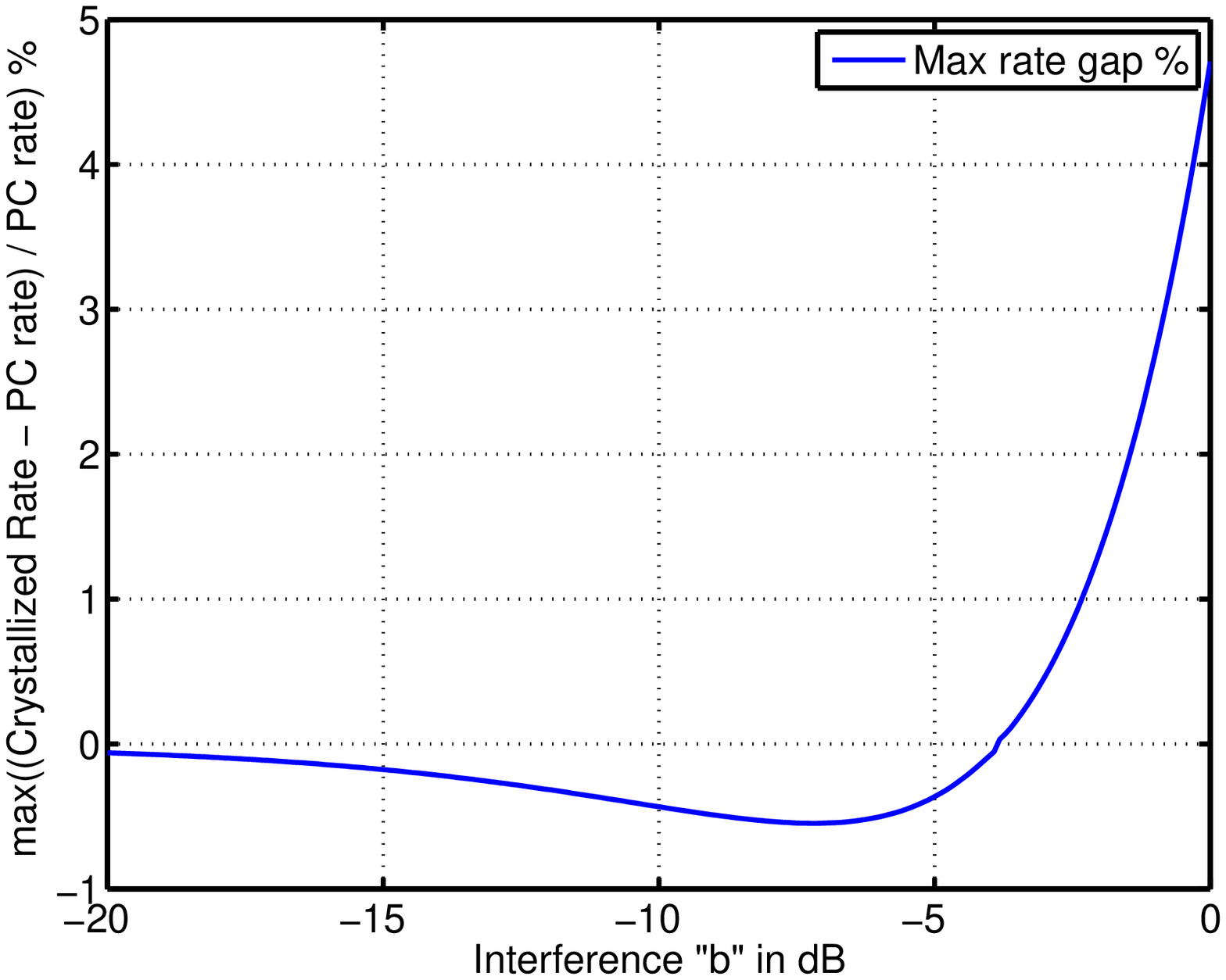}
\caption{(a) comparison of the areas bounded by 
crystallized hull and power control, (b) maximum rate gap percentage of the
difference between crystallized hull and power control rates, with ``b" from
$-20$dB to $0$dB}
\label{fig:RRAreas}\vspace{-5mm}
\end{figure}

The values of the area bounded by the power control potential lines
and the area bounded by the crystallized rate region are 
plotted in Fig.\ref{fig:RRAreas}(a). For weak interference, or equivalently
noise-limited regime, point B is used in constructing the crystallized region.
As the interference increases beyond a certain threshold level, time-sharing
through point B becomes suboptimal, and time-sharing A-C becomes optimal. The
exact switching point between power control to time-sharing is given in
Eq.(\ref{b_threshold}). In Fig.\ref{fig:RRAreas}(a), this happens at the
intersection of the time-sharing line through point B and the A-C time-sharing
line. As
indicated in Fig.\ref{fig:RRAreas}(a), there is no significant loss in the rate
region area if time-sharing is used universally instead of traditional power
control; in fact in some cases time-sharing offers considerable gain.
Specifically, whenever the potential lines exhibit concavity, time-sharing
loses to power control; whenever the potential lines exhibit convexity,
time-sharing wins over power control. Different values of $a$ also lead to the
same conclusion.

The second measure, the maximum rate gap percentage between the rate
achieved by traditional power control and the rate achieved by using the
crystallized rate region is plotted in Fig.\ref{fig:RRAreas}(b). The maximum bit
rate loss of using the crystallized hull compared to power control does not
exceed $1\%$, and the crystallized strategy is therefore quite attractive. 
It is arguable that as the interference becomes the network bottleneck there is
little reason to implement non-binary power control, and that dimension
orthogonalization becomes the primary objective. 

\section{Conclusion}\label{sec:conclusions}
The achievable rate region for the $n-$user interference
channel was presented when the interference is treated as noise. 
The results were first found for the $2-$user interference channel, then they
were extended to the $3-$user case to show the effect of adding an additional
dimension. Subsequently, they were generalized for the $n-$user case. The
$n-$user rate region was found to be the convex hull of the union of the $n$
rate regions, where each rate region is upperbounded by a $(n-1)$-dimensional
hyper-surface characterized by having one of the transmitters transmitting at
full power. For the $2-$user interference channel, those hyper-surfaces become
what the paper refers to as power control potential lines. Hence, to determine
when time-sharing should be used in forming the achievable $2-$user rate
region, the paper studied the potential lines convexity behavior. Finally, the
novel concept of the crystallized rate region was introduced and evaluated. The
crystallized rate region is described by composing a time-sharing convex hull
onto the $2^n-1$ corner points obtained from On/Off binary power control. The
evaluation of the crystallized rate region shows little value in the
implementation of non-binary power control compared to the use of binary power
control in conjunction with time-sharing. In effect,
the crystallized rate region approach offers a new perspective of looking at the
achievable rate region of the interference channel.

\begin{appendices}

\section{Proof that $R_2(P_2)$ is Monotonically
Increasing in $P_2$}
\label{proof_lemma_mono}
\begin{IEEEproof}
Effectively, Eq.~(\ref{C2P2}) is in the form of $f(1+g(x))$. As
$f(\cdot)$ is monotonically increasing in its argument, it suffices
to prove that $g(x)$ is monotonically increasing in $x$. Therefore,
define $g(P_2)$ as:
\begin{equation}
g(P_2) = \frac{acP_2}{a+d(1+bP_2)(2^{r_1}-1)}, 
\end{equation}
\begin{eqnarray}
\frac{\partial g(P_2)}{\partial P_2} 
 = \frac{ac(a+d(2^{r_1}-1))}{(a+d(1+bP_2)(2^{r_1}-1))^2}. \label{del_g}
\end{eqnarray}
The numerator in Eq.~(\ref{del_g}) is nonzero if $a \neq 0$ and $c \neq
0$ ($a=0$ or $c=0$ are the trivial cases where the rate region is
either a line or the point zero). As ${r_1}\geq 0$, then $(2^{r_1}-1)
\geq0$. Thus $\partial g(P_2)/{\partial P_2}$ is always $> 0$ for
non-trivial cases of $a$ and $c$. Thus, $g(P_2)$ is monotonically
increasing in $P_2$, and equivalently $R_2(P_2)$ is monotonically
increasing in $P_2$.
\end{IEEEproof}

\section{Proof that Eq.~(\ref{b_threshold}) is a Sufficient Condition for Both
$\Phi_1$ and $\Phi_2$ to be Convex}\label{App2}
\begin{IEEEproof}
For the symmetric case, $Q_1=Q_2=Q_{sym}$ can be written as
\begin{align}
Q_{sym}
=\frac{\Re(\sqrt{(a-\theta)(a-\theta+a^2P_{\max})})-\theta}{ab},
\nonumber
\end{align}
where $\theta=b+b^2P_{\max}$. $Q_{sym}$ can also be written in this
form:
\begin{align}
Q_{sym} =\frac{\Re(\sqrt{T_1T_2})-\theta}{ab}, \nonumber
\end{align}
where $T_1=a-\theta=a-b-b^2P_{\max}$, and
$T_2=a-\theta+a^2P_{\max}$. From the expression in Eq.~(\ref{b_threshold}), $a$
can be alternatively upper-bounded as $a\leq
(b^2P_{\max}-1/P_{\max})$. Therefore, $T_1$ is upper-bounded as:
\begin{align}
T_1 \leq -1/P_{\max} - b \label{T1}.
\end{align}
From Eq.~(\ref{T1}), $T_1$ is always negative. $T_2$ however can be
positive or negative, evaluated as follows:
\begin{itemize}
\item $T_2\geq 0$: then $\Re(\sqrt{T_1T_2})=0$, and
as $\theta$ is always positive, then $Q_{sym}\leq 0$.
\item $T_2\leq 0$: $\Re(\sqrt{T_1T_2})$ is $ \geq 0$. In this case,
the numerator of $Q_{sym}$ can be written as:
\begin{align}
\mbox{num}(Q_{sym}) =
\sqrt{(\theta-a)(\theta-a-a^2P_{\max})}-\theta. \nonumber
\end{align}
Given the fact that $(\theta-a-a^2P_{\max}) \leq (\theta-a)$, then
$\mbox{num}(Q_{sym})$ can be upper-bounded as:
\begin{align}
\begin{array}{ll}
\mbox{num}(Q_{sym}) & \leq \sqrt{(\theta-a)^2}-\theta  \leq -a \leq 0.
\end{array}\nonumber
\end{align}
\end{itemize}
Hence, the frontiers potential lines $\Phi_1$ and $\Phi_2$ are convex.
\end{IEEEproof}

\section{Extension of Corollary \ref{cor:tdmOptimalSym} to the
$n-$user case}
\label{App3}
Focusing on the $(R_1, R_2)$ 2D rate region of an $n-$user symmetrical rate
region, if Eq.~(\ref{b_threshold}) is satisfied, Appendix~\ref{App2} proved that
it is a sufficient condition to make sure that $\Phi_1$ and $\Phi_2$ are convex
and that they do not have an inflection point. Let $b^*_2$ denote the threshold
value in Eq.~(\ref{b_threshold}). Hence, if the interference channel gain $b \geq b^*_2$,
then $\Phi_1$ and $\Phi_2$ in the $(R_1, R_2)$ 2D rate region are convex. As
more users transmit, they will cause additional interference power on $R_1$ and
$R_2$. Taking the $3-$user case as an example, if the $3^{rd}$ user transmits,
then $R_1(P_{\max},P_2,P_3) = \log_2(1+\frac{aP_{\max}}{1+bP_2+bP_3})$.
Projecting the 3D rate region into the $(R_1, R_2)$ 2D plane will lead to
$R_1(P_{\max},P_2) = \log_2(1+\frac{aP_{\max}}{1+b'P_2})$, where $b' = b +
bP_3/P_2$, which is greater than $b$. Thus, if the $(R_1, R_2)$ 2D rate region
have a convex $\Phi_1$ and $\Phi_2$, then $b \geq b^*_2$; and projecting higher
$n-$dimensional rate region frontiers into the $(R_1, R_2)$ 2D plane will
always result in power control frontiers, denoted as $\Phi_1'$ and $\Phi_2'$,
that are always convex -- due to the fact that they can be obtained using an
interference power gain $b'$ where $b' \geq b \geq b^*_2$.

The goal of this appendix section is to find the $n-$user symmetrical
interference channel gain threshold, denoted as $b^*_n$, such that when $b\geq
b^*_n$, TDM is optimal in leading higher achievable rate region. Thus, as a
starting assumption (which is proved to hold later on), assume $b^*_n
\geq b^*_2, \forall n \geq 2$; which therefore leads to the property that all
the $n-$dimensional hypersurfaces are convex -- as the projections on all the
pair-wise 2D planes result in convex power control frontiers.

With a TDM solution, users transmissions are orthogonal in time, and therefore,
whenever user $i$ transmits all other users are silent. The maximum TDM
rate achievable for user $i$ is when $P_i=P_{\max}$, and in the context of an
$n-$user symmetrical channel, it is equal to $R_i = R^{TDM}_{\cal S}=
\log_2(1+aP_{\max})$, and $R_j = 0, \forall j \neq i$. Therefore, whenever user $i$ transmits under TDM, its
maximum achievable rate in the $n-$dimensional rate region is a point on the
$i^{th}$ axis with a value equals to $R^{TDM}_{\cal S}$; let $\Phi^{TDM}_i$
denotes such a point. 

Let ${\cal H}$ be a hyperplane formed by connecting via
time-sharing the $\Phi^{TDM}_i, i=1,\cdots,n$ points. Define the origin point O
as the point with coordinates $R^O_i=0, \forall i$. Define point B on the
power control achievable rate region when all users transmits at
the same time; thus point B coordinates are $R^B_i =
\log_2\left(1+\frac{\displaystyle
aP_{\max}}{\displaystyle 1+(n-1)bP_{\max}}\right), \forall i$. Define point B'
on the time-sharing hyperplane ${\cal H}$ when all users transmits in TDM for an
equal amount of time. Let $\bf OB$ denotes a vector from point O to point B, and
similarly, $\bf OB'$ denotes a vector from point O to point B'. Let
$||\cdot||$ denotes the vector length; for instance, for $\bf
OB$, $||{\bf OB}||$ is the distance
from point O to point B. Therefore, a time-sharing TDM solution is optimal when
$||{\bf OB'}|| \geq ||{\bf OB}||$; which has the interpretation of
the hyperplane $\cal H$ leading higher achievable rate region than the power
control hyper-surfaces frontiers.

Based on the aforementioned coordinates of point B, $||{\bf OB}|| =
\sqrt{n}\log_2(1+\frac{aP_{\max}}{1+(n-1)bP_{\max}})$. For $||{\bf
OB'}||$, it is the shortest distance from point O to the hyperplane $\cal H$;
which is equal to the absolute value of the dot product of the unit normal
vector of ${\cal H}$ and the vector $\bf OK$, where point K is a point on
${\cal H}$. The unit normal vector of ${\cal H}$ is equal
to $\frac{1}{\sqrt{n}}[1, \cdots, 1]^T$. By choosing point K to be equal to
the point $\Phi^{TDM}_1$ on the $R_1$ axis, $||O\Phi^{TDM}_1|| = [R^{TDM}_{\cal
S}, 0, \cdots, 0]^T$, where $R^{TDM}_{\cal S}$ has been defined earlier to be
equal to $\log_2(1+aP_{\max})$. Therefore, $||{\bf
OB'}|| = \frac{1}{\sqrt{n}}\log_2(1+aP_{\max})$. 

Hence, a time-sharing TDM solution is optimal when 
\begin{align} 
\frac{1}{\sqrt{n}}\log_2(1+aP_{\max}) \geq
\sqrt{n}\log_2\left(1+\frac{aP_{\max}}{1+(n-1)bP_{\max}}\right).
\label{TDM_n_cond}
\end{align}
Expanding Eq.~(\ref{TDM_n_cond}) leads to Eq.~(\ref{b_threshold_n}). The
threshold $b^*_n=\left(\frac{ aP_{\max}}{(1+aP_{\max})^{1/n}-1}-1 \right)\frac{
1}{ (n-1)P_{\max}}$ is monotonically increasing for a
positive $n$, which starts at Eq.~(\ref{b_threshold}) for $n=2$ and flattens out
asymptotically for large n. Using the first order approximation of
$(1+aP_{\max})^{1/n}$ to $1 + \frac{1}{n}\ln(1+aP_{\max})$, $b^*_n$
asymptotically converge to:
\begin{align}
\lim_{n\to\infty}
b^*_{n} = a/\ln(1+aP_{\max}),
\nonumber
\end{align} independent of the number of users $n$.

\end{appendices}

\bibliographystyle{IEEEtran}
\bibliography{bib_entryM}

\begin{biography}[{\includegraphics[width=1in,height
=1.25in,clip,keepaspectratio]{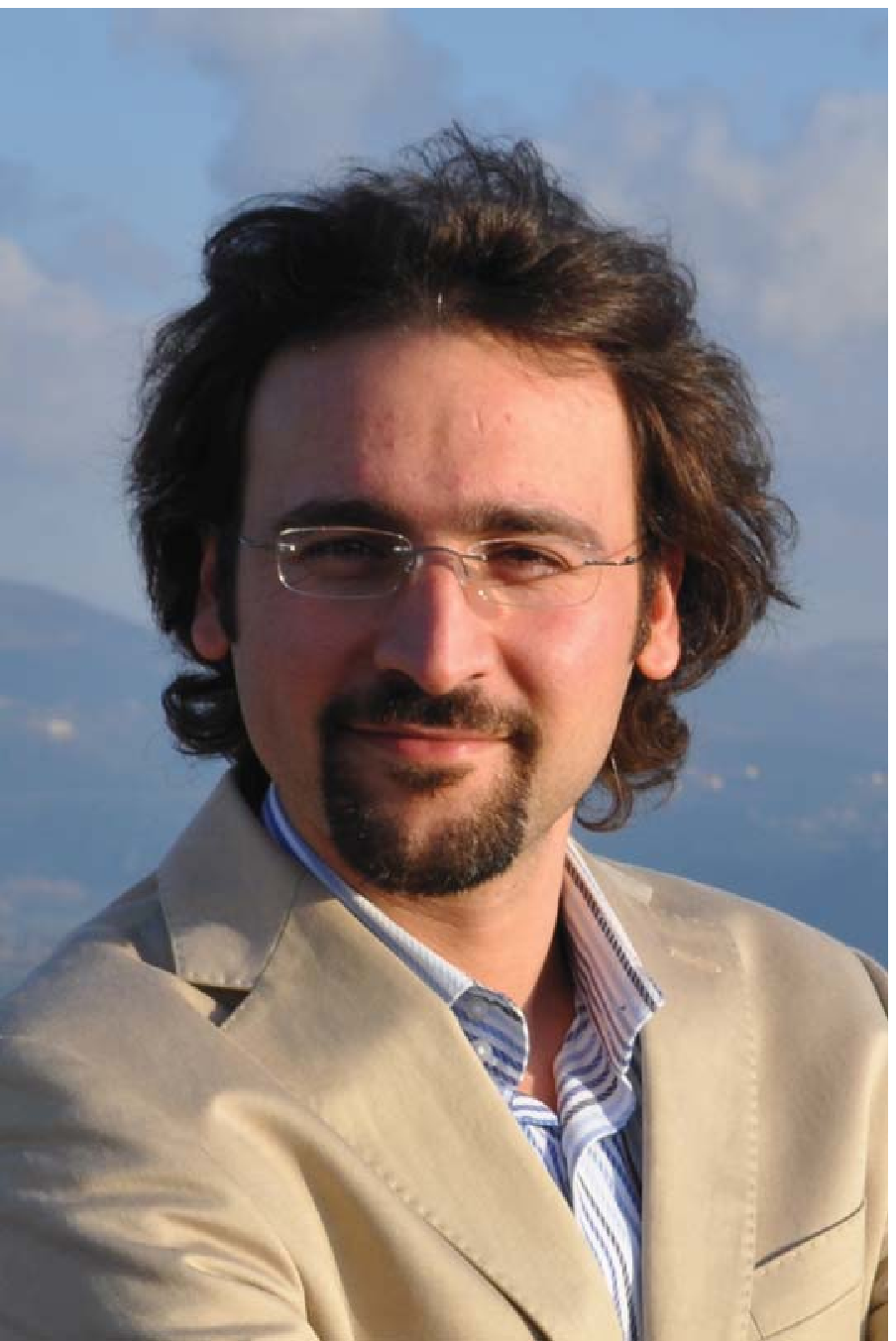}}]
{Mohamad A. Charafeddine \textnormal{received the B.E. in Electrical Engineering
with High Distinction from the American University of Beirut, Lebanon in 2000.
He received his M.S. and Ph.D. in Electrical Engineering from Stanford
University, CA, USA, in 2002 and 2008, respectively. In 2003, he joined and
co-founded two startups in San Diego, CA, USA, and in Amman, Jordan. Since
2008, he joined ASSIA Inc in Redwood City, CA, USA, a software startup
specializing in Dynamic Spectrum Management (DSM), as one of the core
algorithms engineers to manage, diagnose, and optimize DSL networks. In 3 years
since the commercial product launch, ASSIA manages 15\% of the world's DSL
lines. He won a Best Paper Award from the IEEE International Conference on
Communications in 2009.}}
\end{biography}

\begin{biography}[{\includegraphics[width=1in,height
=1.25in,clip,keepaspectratio]{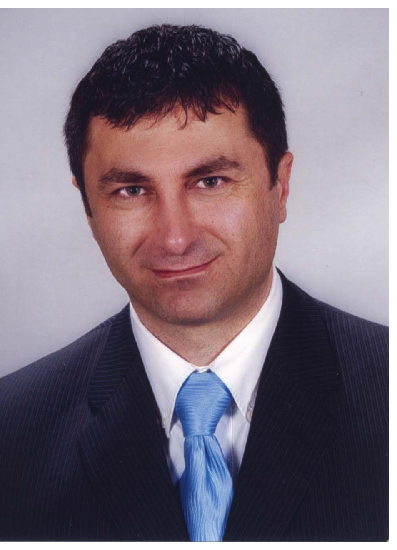}}]
{Aydin Sezgin \textnormal{(S'01 - M'05) received the Dipl.-Ing. (M.S.) degree
in communications engineering and the Dr.-Ing. (Ph.D.) degree in electrical
engineering from the TFH Berlin in 2000 and the  TU  Berlin,  in  2005,
respectively. From 2001 to 2006, he was with the Heinrich-Hertz-Institut (HHI),
Berlin. From 2006 to 2008, he was a Post-doc and Lecturer at the Information
Systems Laboratory, Department of Electrical Engineering, Stanford University.
From 2008 to 2009, he was a Post-doc at the Department of Electrical
Engineering and Computer Science at the University of California Irvine. From
2009 2011, he was the Head of the Emmy-Noether-Research Group on Wireless
Networks at the Ulm University. In 2011, he was full professor at the
Department of Electrical Engineering and Information Technology at TU
Darmstadt, Germany. He is currently a full professor at the Department of
Electrical Engineering and Information Technology at Ruhr-University Bochum,
Germany. His current research interests are in the area of information theory,
communication theory, and signal processing with focus on applications to
wireless communication systems.
He is currently serving as Editor for IEEE
Transactions on Wireless Communications and Area Editor for Elsevier Journal of
Electronics and Communications.}}
\end{biography}

\begin{biography}[{\includegraphics[width=1in,height
=1.25in,clip,keepaspectratio]{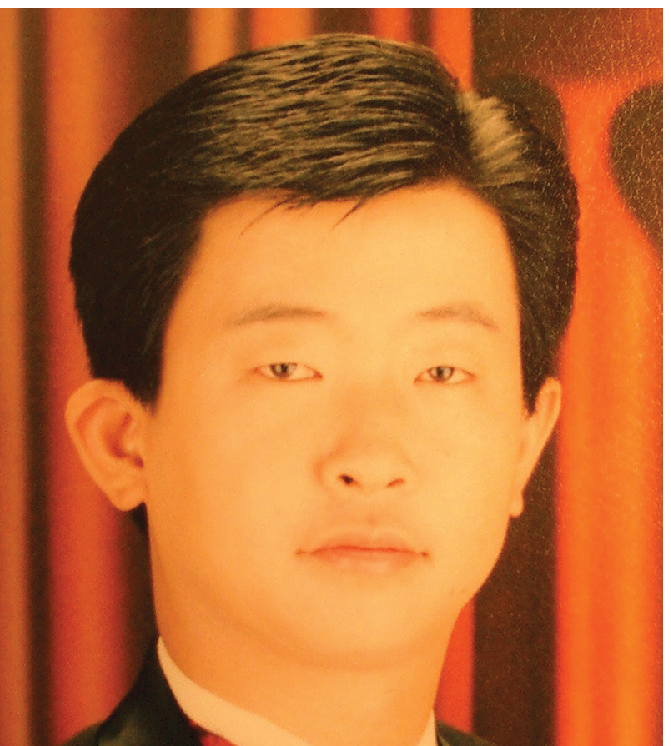}}]
{Zhu Han \textnormal{(S'01 - M'04 - SM'09) received the B.S. degree in
electronic engineering from Tsinghua University, in 1997, and the M.S. and Ph.D. degrees
in electrical engineering from the University of Maryland, College Park, in
1999 and 2003, respectively.
From 2000 to 2002, he was an R\&D Engineer of JDSU, Germantown, Maryland. From
2003 to 2006, he was a Research Associate at the University of Maryland. From
2006 to 2008, he was an assistant professor in Boise State University, Idaho.
Currently, he is an Assistant Professor in Electrical and Computer Engineering
Department at University of Houston, Texas. His research interests include
wireless resource allocation and management, wireless communications and
networking, game theory, wireless multimedia, security, and smart grid
communication.
Dr. Han is an Associate Editor of IEEE Transactions on Wireless
Communications since 2010. Dr. Han is the winner of Fred W. Ellersick Prize 2011. Dr. Han is
an NSF CAREER award recipient 2010. Dr. Han is the coauthor for the papers that
won the best paper awards in IEEE International Conference on Communications
2009 and 7th International Symposium on Modeling and Optimization in Mobile, Ad
Hoc, and Wireless Networks (WiOpt09).}}
\end{biography}

\begin{biography}[{\includegraphics[width=1in,height
=1.25in,clip,keepaspectratio]{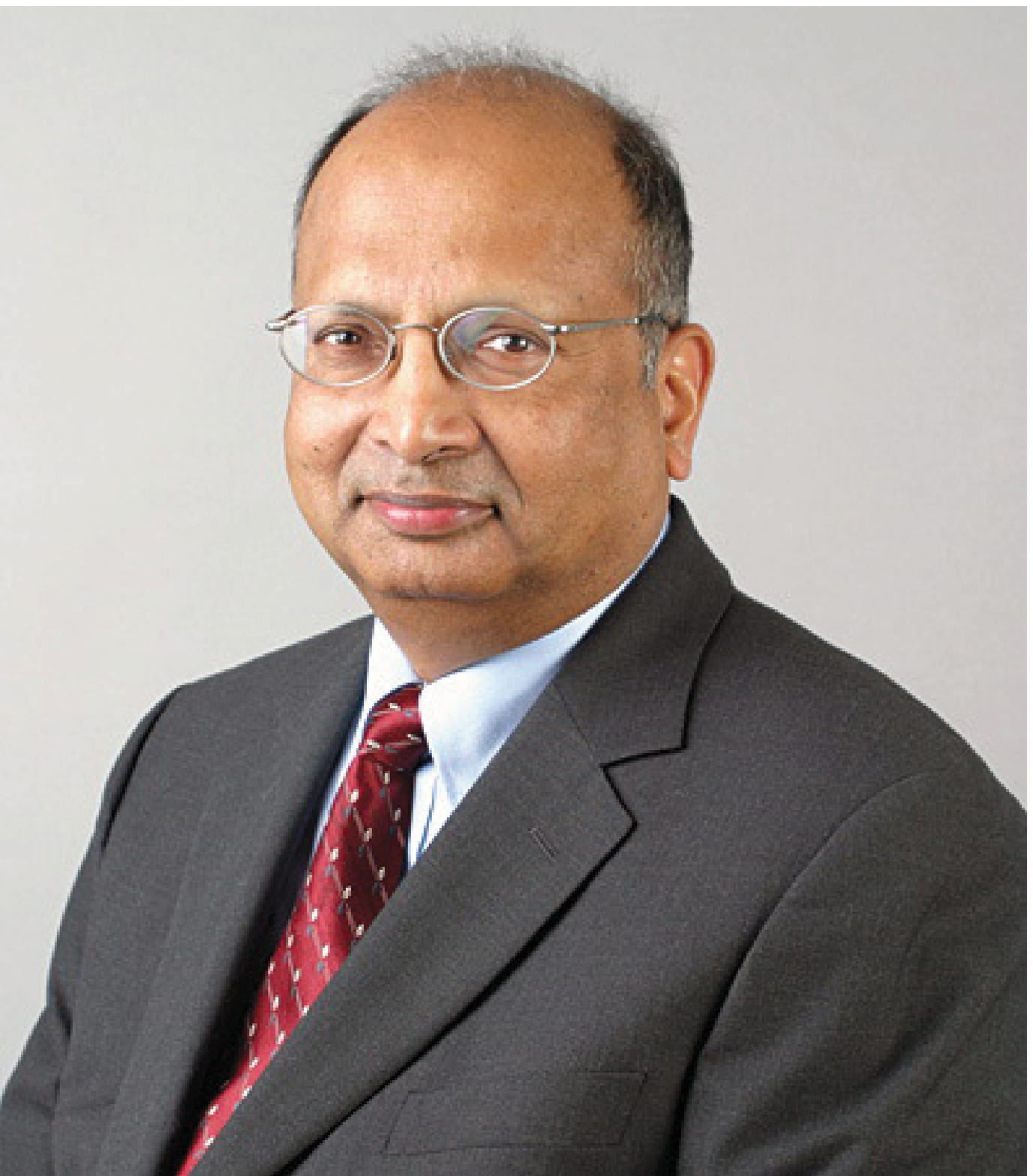}}]
{Arogyaswami J. Paulraj \textnormal{(SM'85 - F'91) is a Professor Emeritus
at Stanford University and the author of over 400 research papers, two
text books and a co-inventor in 52 US patents. He is a fellow of IEEE.
Paulraj has won several awards in the US, notably the IEEE Alexander Graham
Bell Medal. He is a fellow of several scientific academies including the US
National Academy of Engineering.
In 1999, Paulraj founded Iospan Wireless Inc - which developed and established
MIMO/OFDMA wireless as the core 4G technology and was acquired in by Intel
Corporation in 2003. In 2004, Paulraj co-founded Beceem Communications Inc. The
company became the market leader in 4G-WiMAX semiconductor and was acquired by Broadcom Corp. in 2010.
During his 30 years in the Indian (Navy) (1961-1991), he founded three national
level laboratories in India and headed one of India's most successful military
R\&D projects - APSOH sonar. He received many awards in India including the
Padma Bhushan which is one the highest national awards.}}
\end{biography}


%
%


\end{document}